\documentclass[11pt]{article}
\usepackage[utf8]{inputenc}
\usepackage{amsmath, amssymb}
\usepackage{accents}
\usepackage[mathscr]{euscript}
\usepackage{mathrsfs}
\usepackage{mathtools}
\usepackage{textcomp, gensymb}
\usepackage{booktabs}
\usepackage{multirow,makecell}
\usepackage{array}
\usepackage{algorithm, algpseudocode}
\usepackage{comment}
\usepackage{xcolor}
\usepackage{hyperref}
\usepackage[capitalize]{cleveref}
\usepackage{subcaption}
\usepackage[margin=1in]{geometry}

\newcommand\elb[1]{^{\left\{#1\right\}}}

\hypersetup{%
  pdfauthor={Audrey Blizard and Stephanie Stockar},
  pdftitle={Sensitivity Analysis of Performance-Based Partitioning in District Heating Networks},
}

\title{Sensitivity Analysis of Performance-Based Partitioning in District Heating Networks}

\author{
Audrey Blizard\thanks{Department of Mechanical Engineering, The Ohio State University, Columbus, OH 43210. Email: blizard.1@osu.edu}
\and
Stephanie Stockar\thanks{Department of Mechanical Engineering, The Ohio State University, Columbus, OH 43210. Email: stockar.1@osu.edu}
}

\date{}

\begin{document}

\maketitle

\begin{abstract}
\end{abstract}

\noindent\textbf{Keywords:} District Heating Network, Partitioning, Distributed Control, Sensitivity

\begin{abstract}
The paper presents a sensitivity analysis of the factors affecting the optimal partitioning of a district heating network for distributed control. Leveraging a physics-based, distributed model predictive control framework and a performance-based partitioning method, this work studies the relationship between variations in system parameters and the resulting optimal partition, providing insight into the robustness of a nominally designed partition to perturbed operating conditions. The enabling methodology is a learning-enhanced branch and bound method that culls the search space, reducing the number of partitions evaluated for each case. The sensitivity of the nominally optimal partition is characterized across twelve parameter variations, including supply temperature, operating season, building flexibility, pipe characteristics, and building type. This simulation study shows that a well-designed nominal partition exhibits an average cost increase of only 2.8\% relative to centralized control across eleven of the twelve cases, with three cases identifying the nominal partition as globally optimal under the perturbed conditions. The robustness study is followed by an analysis of the sensitivity of the optimality loss metric (OLM), revealing that, in five of twelve cases, the case-specific OLM-minimizing partitions underperform the nominally optimal one due to shifts in the relative magnitude of heat loss versus flexibility costs. This indicates that proper tuning of cost function weights and initial conditions for the performance optimization problem is essential for reliable partition selection, and that seasonal repartitioning is warranted when demand profiles deviate substantially from the nominal, as observed in the November operating case.
\end{abstract}
\maketitle

\section{Introduction}
\noindent The increasing scale and complexity of modern infrastructure systems, such as power grids and district heating networks (DHNs), have made centralized control strategies less practical, driving the need for distributed control approaches that can manage a large number of coupled system elements. Distributed model predictive control (dMPC) has emerged as an effective approach for managing large-scale, nonlinear, dynamic systems like DHNs, due to their computational advantages and ability to incorporate system forecasts. Communication-based dMPC has proven especially useful, allowing the decomposition of the centralized optimization problem into computationally tractable subproblems while maintaining coordination between agents through communication. A key design choice in this framework is how the network is partitioned into subsystems, as this partitioning determines both the computational complexity of the local sub-problem and what information must be communicated, potentially leading to a performance gap between the distributed and centralized controllers. This paper addresses this challenge by conducting a systematic sensitivity analysis of a performance-based partitioning framework, characterizing how the optimal partition and the robustness of a nominally designed one respond to variations in system parameters.\par
Performance-based partitioning methods, such as the optimality loss minimization (OLM)-based framework developed by the authors  \cite{blizardOptimalityLossMinimization2025b}, offer a principled way to select partitions by directly minimizing this performance gap. 
These methods have been particularly effective for systems with nonlinear dynamics, network-wide equality constraints, and complex indirect coupling between component responses, such as DHNs. They have been shown to be near equal in performance to centralized control methods. For a four-user network under nominal operating conditions, the OLM-minimizing partition has been shown to achieve near-centralized performance, with only a 1.7\% increase in cost relative to the centralized controller \cite{blizardOptimalityLossMinimization2025b}. However, this method requires selecting the external inputs before evaluating the partition performance, guaranteeing near-optimal performance only when the controller operates around its nominal design point. Although the found partition can still be used if the nominal scenario is perturbed, the resulting performance may be affected \cite{pourkargarComprehensiveStudyDecomposition2018}.\par
DHNs operate across a wide range of conditions, with heating load, ambient temperature, and supply profiles varying significantly between seasons and times of day. They can also be subject to structural changes, such as changes to the connected buildings and pipes. This variation is known to cause significant changes in the optimal topological design of DHNs. If the design phase considers both summer and winter conditions in an energy-aware manner, the resulting optimal network topology will be very different from one based on a worst-case scenario approach \cite{wackMultiperiodTopologyDesign2024}. While the sensitivity of optimal DHN topology to operating conditions has been characterized, the sensitivity of the optimal partition for control and the robustness of a nominally designed one have not been systematically studied.\par
The need to consider operating conditions during the partitioning process is a recognized challenge, with existing approaches reflecting fundamentally different strategies. For example, in the method using Shapley values to design communication links in the system, large amounts of data are collected to create a probability distribution curve of the expected performance effect of each link \cite{murosGameTheoreticalRandomized2018}, choosing the partition that will perform best on average. The time-averaged RGA was developed to find partitions that are effective for a specific time-scale of interest for linear time-invariant systems \cite{tangRelativeTimeaveragedGain2018}, identifying the partition that is best for a specific case study. Finally, adaptive partitioning methods address condition-dependence by redesigning the partition as operating conditions evolve. In one such approach, the system is repartitioned as the expected state values, and therefore the linearization point, change during operation, using spectral community detection on a time-varying state-space model to identify the updated optimal partitioning \cite{ebrahimiAdaptiveDistributedArchitecture2024}. While effective for linear systems, this approach requires re-solving the partitioning problem in real-time and does not characterize how much the optimal partition actually changes between operating points. In this paper, we examine the sensitivity of the OLM-based partitioning framework to a range of system parameters, including supply temperature, operating season, building flexibility, pipe insulation, pipe diameter, and building type. Two key questions are addressed: whether the optimal partition changes as these parameters vary, and how much performance is sacrificed when a partition designed for nominal conditions is deployed under perturbed ones. This study will enable the designer to decide when the control design should be updated as system conditions evolve.\par
A further challenge in performance-based system partitioning is the $\mathcal{NP}$-hard nature of the element grouping problem \cite{brandesModularityClustering2008}. Because existing partitioning methods do not quantify system-wide performance, but rather the strength of interactions between individual system elements, they naturally reduce to a weighted graph-cutting problem. This structure enables well-established heuristic algorithms, such as modularity maximization and spectral clustering, to find near-optimal solutions efficiently by minimizing the total weight of edges cut between subsystems \cite{jogwarDistributedControlArchitecture2019,ocampo-martinezPartitioningApproachOriented2011}. Performance-based methods, however, do not admit a natural mapping from closed-loop performance degradation to individual edges in a graph. Without this mapping, the search space cannot be pruned using edge-weight-based heuristics, and the quality of a candidate partition can only be assessed by fully evaluating the distributed controller. This means that finding the true OLM-minimizing solution is impractical as the system scale increases. This paper overcomes this challenge using a prescreening method that predicts partition quality without relying on a complete evaluation of the resulting distributed controller. It achieves this by learning the relationship between the communication structure of a partition and its likelihood of convergence, using the existence of communication links between elements as features to predict whether a candidate partition will reach a stable Nash Equilibrium \cite{blizardAcceleratingDistributedControl2025}. By iteratively updating this prediction as new partitions are evaluated, this learning-based approach enables effective prescreening of the search space without requiring full controller evaluation for every candidate.\par
The remainder of the paper is organized as follows. First, \cref{sec:control} presents the specific distributed control framework for large-scale DHNs being considered in this paper. Then, \cref{sec:bnb} presents the learning-based heuristic that enables efficient solving of the optimal partitioning problem. \Cref{sec:results} presents the sensitivity analysis, detailing the problem setup, cases studied, and key results. Finally, \cref{sec:conclusion} presents conclusions and directions for future work.
\section{Distributed Control Design}\label{sec:control}
The control problem considered in this paper is the energy-optimal operation of a district heating network via the heat delivered to individual users. These networks have four types of elements: the feeding pipes, return pipes, bypass pipes, and users. The feeding pipes supply water from the heating plant to the users, while the return pipes circulate the water back to the plant to be reheated. Bypass pipes directly connect these pipe sets, without sending the flow through any buildings. The users connected to the network consist of buildings with integrated heat exchangers to extract heat from the network and a valve to control flow into the heat exchanger. The control variables in the network-wide problem are the mass flow rate supplied by the heating plant, assuming a set supply temperature, and the valve positions for each user. The controller relies on flexibility in the heating demand of the users, where deviations from the nominal demand cause allowable temperature changes in the building. The control developed in this section is a communication-based distributed controller, relying on iterative communication between subsystem controllers to share anticipated variable values. Each local subsystem has three sets of neighbors $\left\{\mathcal{N}\elb{i}_{T}, \mathcal{N}\elb{i}_{\dot{m}}, \mathcal{N}\elb{i}_{P}\right\}$, which communicate their anticipated temperature, mass flow, and node pressure values, respectively, to the subsystem. 
\paragraph{Notation} The communicated value of a variable $x$ is denoted as $\overline{x}$, while $\widetilde{x}$ indicates a vector containing both local and communicated variable values. Additionally, the superscript $\left\{i\right\}$ denotes the subset of global variables local to subsystem $i$. Finally, the subscript "e" indicates all elements, while the subscript "p" indicates non-user edges, the subscript "b" indicates building elements, and the subscript "S" indicates split nodes in the network, where multiple elements are connected. The operation $\left|X\right|$ is the cardinality of set $X$. 
\subsection{Graph Representation}
\begin{figure}\label{fig:graphs}
    \centering
    \begin{subfigure}[t]{.75\linewidth}
        \centering
        \includegraphics[width=\linewidth,clip]{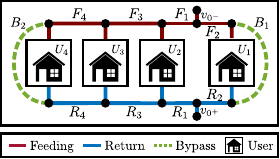}
        \caption{Flow graph.}
        \label{fig:flow_graph}
    \end{subfigure}
    \begin{subfigure}[t]{.75\linewidth}
        \centering
        \includegraphics[width=\linewidth,clip]{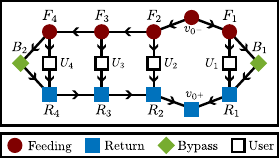}
        \caption{Component connection graph.}
        \label{fig:line_graph}
    \end{subfigure}
    \caption{Modeling graphs for a four-user, mixed parallel series network.}
\end{figure}

The topology of a DHN is represented by two directed, unweighted graphs. The first is the flow graph $\mathcal{G}_{\text{f}} = (\mathcal{V}_{\text{f}},\mathcal{E}_{\text{f}})$ where $\mathcal{E}_{\text{f}}$, the edges, are the network components, and $\mathcal{V}_{\text{f}}$, the nodes, are the flow splits. The plant is modeled as two edges, the input and output, called $\{v_{0^-}, v_{0^+}\}$ respectively. This graph defines the incidence matrix $\Lambda$, used in flow modeling. The second, the component connection graph, $\mathcal{G}= (\mathcal{V}, \mathcal{E})$ is the line graph of $\mathcal{G}_{\text{f}}$ where the nodes $\mathcal{V} =\mathcal{E}_{\text{f}}$ are the network elements and the edges represent heat transfer between these elements. In this graph, the plant is modeled as the root and terminal nodes, $\{v_{0^-}, v_{0^+}\}$, respectively. The non-plant nodes in $\mathcal{G}$ are decomposed into four disjoint sets as $\mathcal{V}=\{F, R,{B},{U}\}$, being the feeding, return, bypass, and user elements, respectively. The graph $\mathcal{G}$ yields an adjacency matrix which identifies the transmission of states between elements. The two graphs associated with a four-user DHN are presented in \cref{fig:graphs}.

\subsection{Local Optimal Control Problem Formulation}
This paper considers a set control formulation, with the design choice being the partitioning of the network into subsystems. Each subsystem solves the local problem given by
\begin{subequations}\label{eq:i_dist}
     \begin{equation}\label{eq:i_cost} 
        \min_{\mathclap{{\dot{m}}_\text{S}\elb{i},\theta_{\text{b}}\elb{i}}}\quad    \sum_{k = t_0}^{t_0+\Delta t} \frac{w_{C}}{\left|{U}\elb{i}\right|}\left(\frac{S_{\text{b}}\elb{i}}{C_{\text{b}}\elb{i}\Delta T_{\text{b}}}\right)^2+w_Q hA\left(T_{\text{p}}\elb{i}-T_{\text{amb}}\right)
    \end{equation}
    \begin{align}
        & \text{subject to:}\notag\\
        \label{eq:i_temp} & T_{\text{p}}\elb{i}(k+1)  = A \left(\dot{m}_{\text{p}}\elb{i}\right) \widetilde{T}_{\text{p}}+ E\left(\dot{m}_{\text{e}}\elb{i}\right)  \begin{bmatrix} T_{0} \\ T_{\text{amb}} \\ T_{\text{setR}} \end{bmatrix}\\
        \label{eq:i_soe} & S_{\text{b}}\elb{i} (k+1) = \dot{Q}_{\text{hx}} -\dot{Q}_{\text{amb}}\elb{i}\\
        \label{eq:i_qp} & \dot{Q}_{\text{hx}} = \dot{m}_{\text{b}}\elb{i} c_p \left(\widetilde{T}_{\text{hx}}-T_{\text{setR}}\right)\\
        \label{eq:i_dP} & \Delta P_{\text{e}}\elb{i} = \zeta_{\text{e}}\elb{i} \left(\dot{m}_\text{e}\elb{i}\right)^2\\
        \label{eq:i_valve} & \zeta_{\text{b}}\elb{i} = \mu\left({\theta_{\text{b}}\elb{i}}^{-1}-1\right)^2\\
        \label{eq:i_pressure} & \Delta P_{\text{e}}\elb{i} = \Lambda_i^\top \widetilde{P}_{\text{S}} \\
        \label{eq:i_flow} & \Lambda_i \dot{m}_{\text{e}}\elb{i} = \widetilde{\dot{m}}_{\text{S}} \\
        \label{eq:i_flex} & -C_{\text{b}}\elb{i} \Delta T_{\text{b}} \leq S_{\text{b}}\elb{i}\leq C_{\text{b}}\elb{i}  \Delta T_{\text{b}}\\
        \label{eq:i_theta} & \theta_{\text{min}}<\theta_{\text{b}}\elb{i}<1,
    \end{align}
\end{subequations}
where the time index $k$ is dropped for compactness. \Cref{eq:i_cost} is the cost to be minimized: a weighted combination of the percent of used flexibility and heat losses to the environment. \Cref{eq:i_temp} determines the temperature dynamics in the pipes. This model takes a bulk temperature approach, where each pipe is a single temperature given by
\begin{equation}
     T_\text{p}\left(k+1\right) = \frac{\dot{m}_\text{p}}{\rho V_\text{p}}\left(T_{\text{in}}\left(k\right) -T_\text{p}\left(k\right) \right)+\frac{hA}{\rho c_p V_\text{p}} \left( T_{\text{amb}}\left(k\right) -T_\text{p}\left(k\right)\right).
\end{equation}
The variable $T_\text{p}$ is the pipe temperature, $T_{\text{in}}$ is temperature of the water flowing into the pipe, $T_{\text{amb}}$ is the ambient temperature, $\dot{m}_\text{p}$ is the mass flow rate through the pipe, and $V_\text{p}, hA, \rho, c_p$ are the characteristics of the pipe and operating fluid. 
Building on this single-pipe equation, the dynamic subnetwork model in \cref{eq:i_temp} is constructed, for all local pipe temperatures $T_\text{p}\elb{i}$. The subset of $\Gamma$ describing the local and neighboring connections is used to assign the values for $T_{\text{in}}$ in the matrix $A\left(\dot{m}_{\text{p}}\elb{i}\right)$, where $T_{\text{in}}$ is either a single pipe temperature in the feeding lines, or a mass-flow-averaged combination of upstream temperatures in the return line. These inlet temperatures are given by
\begin{equation}
    \widetilde{T}_\text{p}= \begin{bmatrix}
        T_\text{p}\elb{i} & \overline{T}_\text{p}\elb{i}
    \end{bmatrix}^\top,
\end{equation}
including both local and communicated pipe temperatures. Finally, the matrix $E\left(\dot{m}_{\text{e}}\elb{i}\right)$ accounts for pipes supplied by the plant, with temperature $T_0$, the heat losses to the environment, and pipes connected to the heat exchangers' outlets, which have an assumed constant temperature, $T_{\text{setR}}$.\par
\Cref{eq:i_soe} determines the local users' states of energy, $S_{\text{b}}$. In this control formulation, the users are considered flexible consumers of heat, where the heat delivered above or below their losses $\dot{Q}_{\text{amb}}$, contributes to their states of energy. The heat supplied by the network, $\dot{Q}_{\text{hx}}$, is given by \cref{eq:i_qp} where $\widetilde{T}_{\text{hx}}\elb{i}\subseteq \widetilde{T}_{\text{p}}$ are the temperatures of the flow into the heat exchangers. \Cref{eq:i_flex} constrains the state of energies based on the buildings' heat storage capacity coefficients $C_{\text{b}}$, a function of the buildings' heat capacities and the acceptable temperature deviation $\Delta T_{\text{b}}$.\par
Finally, \cref{eq:i_dP,eq:i_valve,eq:i_flow,eq:i_pressure,eq:i_theta} are used to calculate the mass flow rate throughout the network. \Cref{eq:i_dP} gives the pressure losses in each local pipe, $\Delta P_\text{p}$, where $\zeta_{\text{e}}$ are the edges' friction coefficients, where the pipes' values, $\zeta_{\text{p}}$, are constant. For the users, the friction coefficients are calculated using \cref{eq:i_valve}, where $\mu$ is a scaling coefficient, and $\theta_{\text{b}}$ is the valve position, constrained by \cref{eq:i_theta} to remain above a minimum pressure loss. \Cref{eq:i_pressure} is the pressure balance equation, ensuring that the pressure drops across all network branches are equal. Here, $\Lambda_i$ is the local subset of $\Lambda$, considering only the local elements, and the flow into and out of those elements, and $\widetilde{P}_\text{S}$ are the split pressures, given by
\begin{equation}\label{eq:constrP}
    \widetilde{P}_\text{S}\elb{v} =\begin{cases}
    \overline{P}_\text{S}\elb{v} & \text{if } v \in\mathcal{N}_{P}\elb{i}\\
    0 & \text{if } v = v_{0^-}\\
    {P}_\text{S}\elb{v} & \text{otherwise}\\
    \end{cases} \quad \forall v \in \mathcal{V}_{\text{f}}\elb{i},
\end{equation}
where the split pressure is associated with its downstream edges in the flow graph for assigning communication edges, and $v_{0^-}$ serves as the reference pressure. \Cref{eq:i_flow} ensures conservation of mass at the splits, where $\widetilde{\dot{m}}_S$ is given by
\begin{equation}
    \widetilde{\dot{m}}_\text{S}\elb{v} =\begin{cases}
    \dot{m}_{\text{S}}\elb{v} & \renewcommand{\arraystretch}{0.9}
        \begin{array}[c]{@{}l@{}}
        \text{if } v = v_{0^-} \\
        \text{or } \exists\, j \neq i \text{ s.t. } v \in \mathcal{N}_{\dot{m}}\elb{j}
        \end{array}\\
    \sum \overline{\dot{m}}_{\text{S}}\elb{v} & \text{if } v\in \mathcal{N}\elb{i}_{\dot{m}}\\
    0 &\text{otherwise}
    \end{cases}\quad \forall v \in \mathcal{V}_{\text{f}}\elb{i}.
\end{equation}
Here $\dot{m}_\text{S}\elb{i}$ is the vector of controllable mass flow rates whose signs match the directed edges of $\mathcal{G}_\text{f}$.\par
Each agent solves this local optimization problem, assuming constant values for $\overline{T}_\text{p}\elb{i}, \overline{P}_\text{S}\elb{i}$, and $\overline{\dot{m}}_{\text{S}}\elb{i}$. 
After each solution iteration, the values from the neighboring optimization problems are sent to the local subsystem and updated according to
\begin{equation}
\label{eq:convergence}
    \begin{bmatrix}
        \overline{T}_\text{p}\elb{i}\\
        \overline{P}_\text{S}\elb{i}\\
        \overline{\dot{m}}_{\text{S}}\elb{i}
    \end{bmatrix}_{n_{\text{iter}}\elb{i}+1}=\omega \begin{bmatrix}
        T_\text{p}\elb{\mathcal{N}_{T}\elb{i}}\\
        P_\text{S}\elb{\mathcal{N}_{P}\elb{i}}\\
        \dot{m}_{\text{S}}\elb{\mathcal{N}_{\dot{m}}\elb{i}}
    \end{bmatrix}_{n_{\text{iter}}\elb{i}} + \left(1-\omega\right)\begin{bmatrix}
        \overline{T}_\text{p}\elb{i}\\
        \overline{P}_\text{S}\elb{i}\\
        \overline{\dot{m}}_{\text{S}}\elb{i}
    \end{bmatrix}_{n_{\text{iter}}\elb{i}},
\end{equation}
with step size $\omega$, where $n_{iter}\elb{i}$ is the solution iteration for the local subsystem. This process is repeated until convergence. The subsystem problem is considered converged when the changes in local variables and cost are less than a vector of threshold values $\epsilon$ and when all neighboring subsystems that pass information into the local system also meet this criterion. Note that initial guesses for the values $\overline{T}_\text{p}\elb{i}, \overline{P}_\text{S}\elb{i}$, and $\overline{\dot{m}}_{\text{S}}\elb{i}$ must be provided for the first iteration.\par
\subsection{Optimal Partition Selection}
The goal in designing this distributed controller is to select the network partitioning that minimizes the performance degradation of the distributed controller as compared to the centrally optimal solution. Partitioning is defined as separating the nodes in the component graph $\mathcal{G}$ into disjoint sets as
\begin{equation}
\begin{split}
    \operatorname{part}\left(\mathcal{V}\right) &= \left\{\mathcal{V}\elb{1}, \dots, \mathcal{V}\elb{n_p} \right\},\ \mathcal{V}\elb{i} \subseteq \mathcal{V}\\
    \text{ s.t. } \mathcal{V} &= \bigcup_{i = 1}^{n_p} \mathcal{V}\elb{i},\quad
    \mathcal{V}\elb{i}\cap \mathcal{V}\elb{j}=\emptyset\ \forall i\neq j,
\end{split}
\end{equation}
where each set $\mathcal{V}\elb{i}$ is a subsystem of the system components to be controlled. Each system component is associated with a temperature, flow, and node pressure value, meaning there are three variable sets, temperature, flow rate, and pressure, to be passed between the created subsystems. These variables are passed in different directions to ensure consistency and allow at least one controllable value per component. The direction of communication is encoded in a communication graph $\mathcal{G}_C = \left(\mathcal{E}_C, \mathcal{V}\right)$, where $\mathcal{E}_C$ has edges to direct each of the three variable sets, $\mathcal{E}_C = \left\{\mathcal{E}_T, \mathcal{E}_{\dot{m}}, \mathcal{E}_{P}\right\}$. Their directions are defined based on the direction of flow in the edge set $\mathcal{E}$ as
\begin{equation}
    \mathcal{E}_T = \mathcal{E}
    \end{equation}
    \begin{align}
        \mathcal{E}_{\dot{m}} & = \left\{ \begin{array}{l} (v_j,v_i) \\ (v_i,v_j) \\ \end{array} \middle| \begin{array}{l} v_i\in F^0,\ v_j\in F^* \\ v_i\in R^*,\ v_j\in R^0 \end{array}, (v_i,v_j)\in\mathcal{E} \right\}\\
        \mathcal{E}_{P} & = \left\{ \begin{array}{l} (v_i,v_j) \\ (v_j,v_i) \\ \end{array} \middle| \begin{array}{l} v_i\in F^0,\ v_j\in F^* \\ v_i\in R^*,\ v_j\in R^0 \end{array},  (v_i,v_j)\in\mathcal{E} \right\}
    \end{align}
using $\left(v_i,v_j\right)$ to indicate the edge from node $v_i$ to $v_j$, and $F^*= F\cup U\cup B$, $R^* = R\cup U\cup B$, $F^0=F\cup v_{0^-}$, and $R^0= R\cup v_{0^+}$.
Once the system is partitioned for distributed control, any intra-subsystem edges represent variables that must be communicated during the distributed control process. The nodes with information that must be communicated to a subsystem are assigned to the subsystem's neighboring set, defined as
\begin{equation}
    \mathcal{N}\elb{i} = \left\{ v_j\ \middle|\ \exists\ \left(v_j,v_i\right)\in \mathcal{E}_C: v_i\in\mathcal{V}\elb{i} \right\}.
\end{equation}
This definition is used to create neighbor sets, $\left\{\mathcal{N}\elb{i}_T,\mathcal{N}\elb{i}_{\dot{m}},\mathcal{N}\elb{i}_P\right\}$, as previously described, corresponding to the three sets of the communication edges.\par
Finding the optimal network partitioning for the described control structure leverages a game-theoretic perspective to minimize the performance degradation of the distributed system. The optimal partitioning problem is defined as
\begin{subequations}\label{eq:olm}
    \begin{equation}
        \min_{\operatorname{part}\left(\mathcal{V}\right)} c_{\text{olm}}
    \end{equation}
    \begin{align}
        & \text{subject to:}\notag\\
        \label{eq:colm} & c_{\text{olm}} = w_{\text{mPoA}}c_{\text{mPoA}}+w_{\text{sz}}c_{\text{sz}}+w_{\text{iter}}c_{\text{iter}}\\
        \label{eq:mpoa} &c_{\text{mPoA}} = \frac{c_{\text{cen}}\left(u_{\text{NEs}}\right)}{c_{\text{cen}}\left(u_{\text{opt}}\right)}\\
        \label{eq:sz} &c_{\text{sz}} = \max_i \left( \left| \mathcal{V}\elb{i}\right| \right)\\
        &c_{\text{iter}} = \max_{i} \left(n_{\text{iter}}\elb{i}\right).
    \end{align}
\end{subequations}
The goal of this optimization problem is to balance performance degradation with computation time, captured by the Optimality Loss Metric (OLM). The OLM defined in  \cref{eq:colm} has three weighted terms. The first is the modified price of anarchy (mPoA). This term, \cref{eq:mpoa}, is the ratio of the cost associated with the found stable Nash Equilibrium (sNE) control action to the cost associated with the true optimal solution. Here, the term "stable" refers to the game-theoretic concept, where the NE consensus is not easily escaped, even with numeric stochasticity or alternate solvers \cite{seatonIntrinsicFragilityPrice2023}. The control input $u_{\text{NEs}}$ is found via \cref{eq:i_dist}, dependent on the network partitioning. 
The second term, given in \cref{eq:sz} is the size of the largest partition, which directly affects the local optimization problem complexity and encourages partitioning beyond the centralized solution. The final term is the iterations needed for the distributed controller to reach a consensus, which is included to select partitions that quickly reach a consensus about their preferred actions. These terms are weighted with decreasing priority by weights $w$. 

\section{Solution Method}\label{sec:bnb}
The optimization problem in \cref{eq:olm} is an $\mathcal{NP}$-hard integer programming problem, and therefore finding the exact solution is impractical for larger networks \cite{shiNormalizedCutsImage2000}.  This section presents a heuristic method, the learning enhanced branch and bound (LE-BnB), that enables the pre-screening of candidate solutions to predict convergence. This method uses the iterative bi-partitioning structure to approximate the true optimal solution. Because the LE-BnB relies on the structure of an exact branch and bound algorithm, it is effective in finding the optimal system partitioning, as long as the recall of the pre-screening step is sufficiently high.\par
Note that for the specific DHN control problem being considered here, $v_{0^+}$ is always assigned to a separate partition, and $v_{0^-}$ is always included in a partition with at least one connected node. This method has been adapted from prior work \cite{blizardAcceleratingDistributedControl2025} to incorporate an active learning component, refining the critical edge learner as solutions are explored, which is key for larger systems that require higher levels of partitioning. 
\begin{algorithm}
    \caption{LE-BnB for graph partitioning.}
    \label{alg:bnb}
    \begin{algorithmic}[1]
    \State{$c_{\text{olm}}(b)\gets \infty$, $\mathcal{P}_{0}\gets \mathcal{V}, \mathcal{P}_{\text{train}}\gets \emptyset$}
    \For{$n=1\dots n_{\text{e}}$}
        \State{Initialize $\mathcal{P}_{n}\gets \emptyset$}
        \For{$i\ =1\dots \left|\mathcal{P}_{n-1}\right|$}
            \State{$ \overline{\mathcal{P}}_{n} \gets$  \Call{Branch}{$\mathcal{P}_{n-1}\elb{i}$}}
            \State{Append $\overline{\mathcal{P}}_{n}$ to $\mathcal{P}_{n}$}
        \EndFor
        \State{$b_{\text{solved}}\gets \mathbf{false}^{\left|\mathcal{P}_{n}\right|}$}
        \If{$n>1$}
            \State{$\mathcal{S}\gets$ \Call{FindCand}{$\mathcal{P}_{n},b_{\text{solved}},\operatorname{CEL}$}}
        \Else
            \State{$\mathcal{S} \gets \operatornamewithlimits{rand}_{n_{\text{train}}}\left(\mathcal{P}_{n}\right)$}
        \EndIf
        \While{$\left|\mathcal{S}\right|>0$}
            \State{$b_{\text{solved}}(\mathcal{S}) =$ true}
            \label{algl:parallelize}\If{$\min_{i}\left(c_{\text{olm}}\left(\mathcal{S}\elb{i}\right)\right)<c_{\text{olm}}\left(b\right)$}
            \State{$b\gets \mathcal{S}\elb{i}$ s.t. $i =\operatorname{argmin}_i \left( c_{\text{olm}}\left( \mathcal{S}\elb{i} \right) \right)$}
            \EndIf
            \State{$\mathcal{P}_{\text{train}}\gets \mathcal{P}_{\text{train}}\cup \mathcal{S}$}
            \If{$\left|\mathcal{S}\right|>n_{\text{rt min}}$}
            \label{algl:trainCEL}\State{$\operatorname{CEL}\gets$ \Call{TrainCEL}{$\mathcal{P}_{\text{train}},b_{\text{conv act}}\left(\mathcal{P}_{\text{train}}\right)$}}
            \EndIf
            \If{$n>1$}
            \State{Remove $\mathcal{P}_{n}\elb{i}$ s.t. $\widetilde{c}_{\text{olm}}\left(\operatorname{parent} \mathcal{P}_{n}\elb{i} \right) > c_{\text{olm}}\left(b\right)$}
            \EndIf
            \State{$\mathcal{S}\gets$ \Call{FindCand}{$\mathcal{P}_{n},b_{\text{solved}},\operatorname{CEL}$}}
        \EndWhile

        \State{$\mathcal{P}_{n}\gets \mathcal{P}_{n}\elb{i} \forall i \text{ s.t. } \widetilde{c}_{\text{olm}}\left(\mathcal{P}_{n}\elb{i}\right)\leq c_{\text{olm}}(b)$}
        \If{$\left|\mathcal{P}_{n}\right|=0$}
        \State{\textbf{break}}
        \EndIf 
    \EndFor
    \State{\textbf{return} $b$}
    \end{algorithmic}
\end{algorithm}
\begin{algorithm}
    \caption{Procedures used in the LE-BnB.}
    \label{alg:bnb_procedure}
    \begin{algorithmic}[1]
    \Procedure{FindCand}{$\mathcal{S},b_{\text{solved}},\operatorname{CEL}$}
    \State{$b_{\text{conv pred}} \gets \operatorname{CEL}\left(\mathcal{G}_A\left(\mathcal{S}\right)\right)$}
    \State{$\overline{S}\gets b_{\text{conv pred}} \land \neg b_{\text{solved}}$}
    \State{$\overline{S}\gets \overline{S}\left(1:\min{\left(\left|\overline{S}\right|, n_{\text{rt}}\right)} \right)$}
    \State{\textbf{return} $\overline{\mathcal{S}}$}
    \EndProcedure
    \par\vskip.2\baselineskip\hrule height .4pt\par\vskip.2\baselineskip
     \Procedure{Branch}{$p$}, $p = \left\{\mathcal{V}\elb{1},\dots,\mathcal{V}\elb{n}\right\}$
    \State{Initialize $\mathcal{S}\gets \emptyset$}
    \If{$\left|\mathcal{V}\elb{n}\right|>1$}
        \State{$\overline{\mathcal{P}}_{2} = \left\{ \overline{p}\ \middle|\ \overline{p} = \operatorname{part} \left(\mathcal{V}\elb{n}\right), \left| \overline{p}\right| = 2\right\}$}
        \For{$i = 1 \dots \left| \overline{\mathcal{P}}_{2} \right|$}
            \State{$\overline{\mathcal{P}}_{2}\elb{i} = \left\{ \overline{\mathcal{V}}_{i}\elb{1},\overline{\mathcal{V}}_{i}\elb{2} \right\}$}
            \State{$p_{\text{new}} = \left\{\mathcal{V}\elb{1}, \dots, \mathcal{V}\elb{n-1}, \overline{\mathcal{V}}_{i}\elb{1},\overline{\mathcal{V}}_{i}\elb{2} \right\}$}
            \If{$p_{\text{new}}$ is valid}\label{algl:check}
                \State{Append $p_{\text{new}}$ to $\mathcal{S}$}
            \EndIf
        \EndFor
    \EndIf 
    \State{\textbf{return} $\mathcal{S}$}
    \EndProcedure
    \end{algorithmic}
\end{algorithm}
\subsection{Critical Edge Learner}
In a full hierarchical BnB search for the optimal partitioning of the DHN in \cref{fig:graphs}, 93.8\% of solutions do not converge to a Nash Equilibrium solution. The Critical Edge Learner (CEL) is used to prescreen the solutions to predict if the partitioned system will converge to a consensus. An effective indicator weather a system converges is each element's awareness of others' behaviors \cite{blizardAcceleratingDistributedControl2025}. This awareness comes from elements in the same subsystem via the local cost function, modeled via a graph, where the existence of an edge between an element indicates awareness, given by $\mathcal{G}_A = \left({\mathcal{V},\mathcal{E}_A}\right)$. When the graph is fully connected, every element is aware of the others' behaviors and effects on the system, as seen in a centralized controller. This fully connected graph has 
\begin{equation}
    \operatorname{card}\left(\mathcal{E}_{A}\right) = \frac{\operatorname{card}\left(\mathcal{E}_{D}\right)\left(\operatorname{card}\left(\mathcal{E}_{D}\right)-1\right)}{2}
\end{equation}
edges, which are the training features for the CEL. As the system is partitioned, edges are removed to separate the elements into their respective subsystems. The existence of these edges is a binary variable used as the training features in the CEL. The chosen training model is a bagged tree classifier, a common choice for binary classification problems \cite{jamesIntroductionStatisticalLearning2013}. A bagged tree classifier determines a classification score by averaging the outcomes of multiple binary classification trees. The chosen training process uses random subsets of the data to construct these binary classification trees called weak learners in the process of bootstrap sampling \cite{jamesIntroductionStatisticalLearning2013}.
\subsection{Learning Enhanced Branch and Bound}
The basis of the LE-BnB is the assumption that further division of a subsystem can only serve to decrease system performance. This assumption allows the search to be structured as a tree where a multi-cut partition is reached through a series of iterative bi-partitions. The tree structure allows partition branches to be trimmed as soon as they have no chance of outperforming the current bounding solution. However, even with this tree structure, there are still at minimum $2^{n_{\text{e}}-1}$ branches to be explored, where $n_{\text{e}}$ is the number of elements in the system, with the maximum number of solutions to explore being Bell's number $B_{n_{\text{e}}}=\sum_{k=0}^{n_{\text{e}}-1}\binom{n-1}{k}B_k$.\par
The LE-BnB uses the CEL as a prescreener to determine which solutions to evaluate the OLM for. First, a random subset of the 1-cut solutions is selected and evaluated for convergence. These are used to train the initial CEL. Then, all potential 1-cut solutions are prescreened, and the first set of $n_{\text{rt}}$ solutions predicted to converge are evaluated. Next, the CEL is retrained, and the process is continued until the number of unexplored predicted-converging solutions is less than the retraining lower limit $n_{\text{rt min}}$. From here, the retraining stops, and the remaining solutions are evaluated.\par
To select solutions for future bi-partitioning, the criteria 
\begin{gather}
    \widetilde{c}_{\text{olm}}(\bar{p}) <c_{\text{olm}}(b)\\
    \widetilde{c}_{\text{olm}} = w_{\text{mPoA}} c_{\text{mPoA}} +w_{\text{sz}} \widetilde{c}_{\text{sz}} +w_{\text{iter}} c_{\text{iter}}
\end{gather}
is used, where $b$ is the current best solution, and $\widetilde{c}_{\text{olm}}$ is an optimistic $c_{\text{olm}}$, using the size of the largest subsystem not to be further subdivided ($\widetilde{c}_{\text{sz}}$), rather than the overall largest subsystem. This process is repeated until all branches have either been eliminated or fully explored. \par
The pseudocode for this process is presented in \cref{alg:bnb}. Note that $\mathcal{P}_n$ is a set of partitions with $n$ agents. Additionally, $\mathcal{P}_{\text{train}}$ is the set of all partitions to be used as training data for the CEL.  This data is used in \cref{algl:trainCEL} to iteratively update the CEL as more solutions are explored. This set can be trimmed as solutions are explored if storage or training time is a limiting factor. In a slight abuse of notation, $\mathcal{S}$ is used both as a subset of partitions and the index of that subset in the overall list. The variable $b$ is used for binary vectors, where $b_{\text{solved}}$ indicates the problem has been previously considered, $b_{\text{conv pred}}$ indicates the convergence predicted by the CEL, and $b_{\text{conv act}}$ indicates the true convergence status of the partitions, after evaluation. The operator $\operatornamewithlimits{rand}_{n}(\mathcal{S})$ is the random selection of $n$ partitions of $\mathcal{S}$. Additionally, \cref{algl:check} ensures that the heating plant remains in a subsystem with one of its directly connected edges, to ensure a determinate plant flow rate. Finding the OLM for each partition in $\mathcal{S}$ in \cref{algl:parallelize} is parallelizable to enable faster solving.\par
\begin{figure}
    \centering
    \includegraphics[width=1\linewidth]{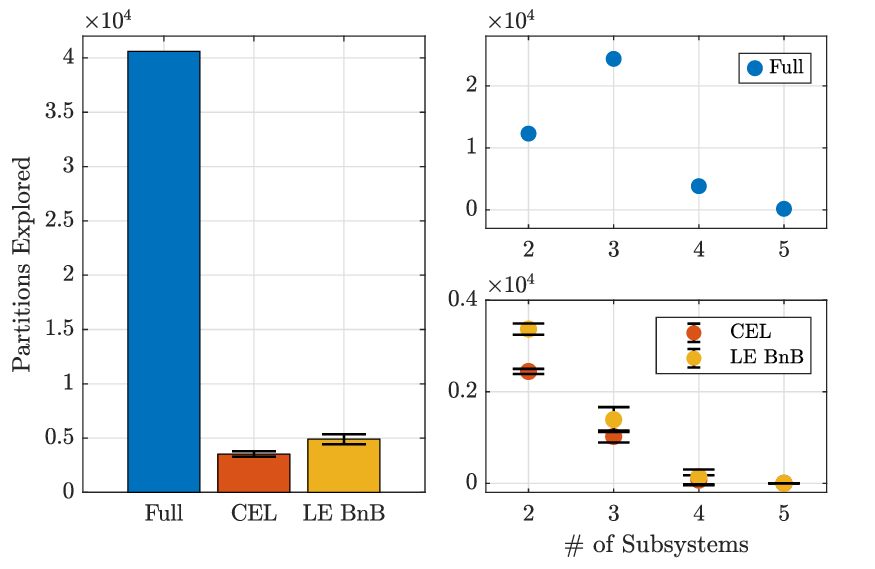}
    \caption{Partitions explored by the complete and learning-enhanced algorithms, with standard deviation over 10 trials.}
    \label{fig:part_explored}
\end{figure}

In this paper, the following parameters were used for the LE-BnB algorithm. The bagged tree classifier used 30 learners. Additionally, to favor identifying all converging partitions, false negatives were penalized four times more than false positives during the training. For the branch and bound algorithm, $n_{\text{train}}= 1000$, $n_{\text{rt}}= 1000$ and $n_{\text{rt min}}= 100$, to balance accuracy and the need to retrain the learner. The efficacy of the LE-BnB is compared to an exact BnB without prescreening and a constant CEL trained on the first 1000 partitions. The validation is performed on the DHN configuration shown in \cref{fig:graphs} with nominal operating conditions. The heuristic search was performed 10 times to account for stochasticity in the initial training set selection. The LE-BnB was able to find the global optimal partitioning in 100\% of evaluations, while the constant CEL only identified the correct partition in 50\% of the trials. Additionally, the LE-BnB only explored on average 4,900 partitions compared to 40,000 in the full search. The number of solutions explored by each algorithm in total and decomposed by number of iterative cuts, with variance over the ten runs, is presented in \cref{fig:part_explored}.

\begin{figure}
    \centering
    \includegraphics[width=.7\linewidth]{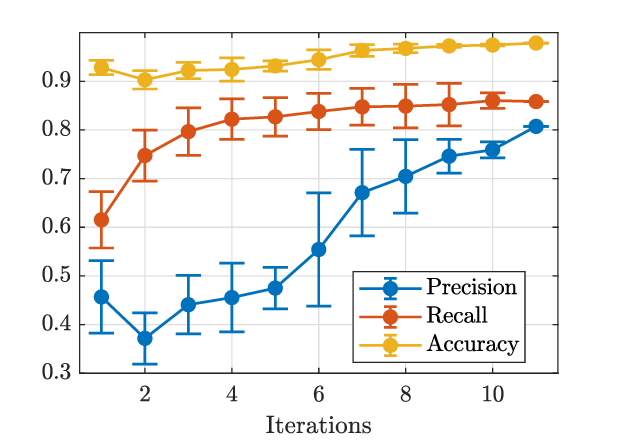}
    \caption{Mean accuracy, precision, and recall of the CEL as the exploration evolves, with STD over 10 trials.}
    \label{fig:cel_metric}
\end{figure}

Additionally, the performance of the CEL over solution iterations is plotted in \cref{fig:cel_metric}. This figure shows the accuracy, precision, and recall of the learner over all partitions evaluated by the full search. Note that the standard deviation in these values decreases at higher iterations because the number of iterations needed to explore the search space varied between trials, so only a few trials have the maximum iterations. The key metric is recall, which quantifies the percentage of converging partitions accurately identified by the CEL, averaging 85\% by the end of the solution iterations.

\begin{table}
\caption{Nominal network parameters.}
\label{tbl:nom_params}
\centering
\begin{tabular}{l c c c}
\toprule
Parameter & Symbol & Value & Units\\
\midrule
Supply temperature & $T_0$ & 80 & $C$\\
Average ambient temperature & $T_{\text{amb}}$ & -14 & $C$\\
Pipe diameter & $D$ & 0.40 & $m$\\
Heat transfer coefficient & $h$ & 1.5 & $W/m^2K$\\
Building flexibility & $\Delta T_b$ & 2 & \degree$C$\\
\bottomrule
\end{tabular}
\end{table}

\section{Sensitivity Analysis}\label{sec:results}
\begin{figure}
    \centering
    \begin{subfigure}[t]{0.85\linewidth}
        \includegraphics[width=\linewidth]{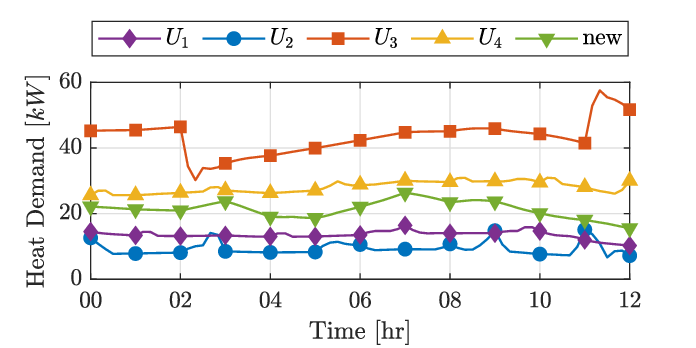}
        \caption{Nominal, plus the new demand used in cases l,m.}
        \label{fig:demand}
    \end{subfigure}
    \\\vspace{10pt}
    \begin{subfigure}[t]{.48\linewidth}
        \includegraphics[width=\linewidth, trim={0 0 0 45}, clip]{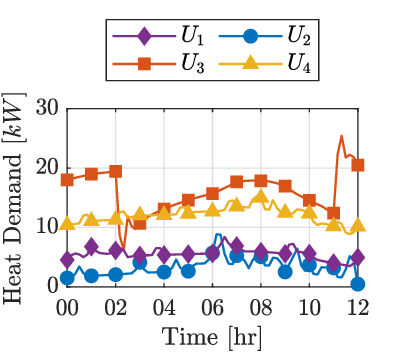}
        \caption{November.}
        \label{fig:demand_nov}
    \end{subfigure}%
    \hfill
    \begin{subfigure}[t]{0.48\linewidth}
        \centering 
        \includegraphics[width=\linewidth, trim={0 0 0 45}, clip]{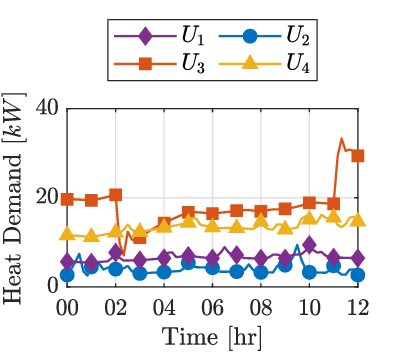}
        \caption{December.}
        \label{fig:demand_dec}
    \end{subfigure}
    \caption{Demands for each operating season of the four buildings.}
    \label{fig:demand_seasonal}
\end{figure}

The objective of this sensitivity analysis is to determine the effects of operational and design changes on the performance of a partition. Six variables were considered: supply temperature, operating season, flexibility limits, building types, pipe diameter, and network insulation. All changes were compared to a nominal case. The conditions for the nominal operating case are summarized \cref{tbl:nom_params}. The nominal operating time period was the first week of January, with building demand, shown in \cref{fig:demand}, taken from NREL \cite{horowitz2019resstock}. The nominal heat capacities of the four buildings in the network and lengths of the network pipes are presented in \cref{tbl:network}.\par

\begin{table}
\caption{Component parameters.}
\label{tbl:network}
\begin{minipage}{0.48\linewidth}
    \centering
    \begin{tabular}{l c }
    \toprule
    $\mathcal{E}_{\text{f}}$ & \makecell{Length\\($m$)}\\
    \midrule
    $F_1,R_1$ & 80 \\
    $F_2,R_2$ &60 \\
    $F_3,F_4,R_3,R_4$ & 20\\
    $B_1,B_2$ & 3\\
    \bottomrule
  \end{tabular}
\end{minipage}
\hfill
\begin{minipage}{0.48\linewidth}
  \centering
  \begin{tabular}{l c c }
    \toprule
    $\mathcal{E}_{\text{f}}$ & \makecell{$C$ \\ ($MJ/K$)} & \makecell{Init. $S_{\text{b}}$\\ (\%)}\\
    \midrule
    $U_1$ & 526 & -4\\
    $U_2$ & 78 & 8\\
    $U_3$ & 400 & -10\\
    $U_4$ & 901 & 2\\
    new & 2390 &  - \\
    \bottomrule
  \end{tabular}
\end{minipage}
\end{table}
\begin{table}
\caption{Parameters for simulation.}
\label{tbl:sim}
\centering
\begin{tabular}{r l l }
\toprule
Equation & Eq. \#& Values \\
\midrule
OLM &\eqref{eq:colm} & $w_{\text{mPoA}}$ = 1, $w_{\text{iter}}$= .04, $w_{\text{sz}}$ = .06  \\
Cost& \eqref{eq:i_cost}& $w_C$ = 5, $w_Q$ = $3\times 10^{-6}$ \\
$\epsilon$& \eqref{eq:convergence} & ${\dot{m}_{\text{p}}}$ = 0.2 $kg/s$, ${\dot{m}_{0}}$= 0.3 $kg/s$,\\
& & ${P_{\text{S}}}$= 5 $kPa$, local cost: 0.5\\
\bottomrule
\end{tabular}
\end{table}
\begin{figure}
    \centering
    \includegraphics[width=.75\linewidth]{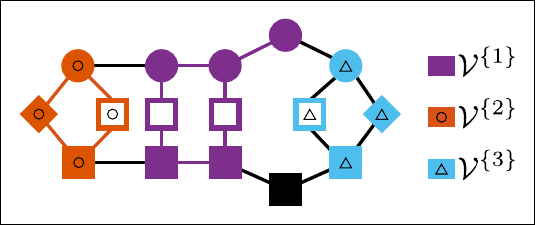}
    \caption{Baseline partition.}
    \label{fig:part_base}
\end{figure}
In all simulations, the friction coefficient is $\zeta= 1\ \left(km\cdot kg\right)^{-1}$, the minimum valve position is $\theta_{\text{min}} = 0.01$, the valve coefficient is $\mu=2.6$, and the bypass diameter $0.15\ m$. The cost function weights are presented in \cref{tbl:sim}. All cases also use the same evaluation and simulation settings. The simulations are performed in discrete time with a 10-minute control time step, a 30-second discretization of temperature variables, a 1-hour optimization horizon, and a 10-minute rollout horizon. The value for $\omega$ in \cref{eq:convergence} is 0.5. The calculation of $c_{\text{olm}}$ is performed using a single optimization step, with the weights in \cref{tbl:sim}. The simulations were performed over a 12-hour window. All optimization was performed using CasADi \cite{Andersson2019} and IPOPT \cite{wachterImplementationInteriorpointFilter2006}.\par 
The costs for each control method in the nominal case are shown in \cref{fig:cost_nom}. The nominal partition in \cref{fig:part_nom} has three agents, with the plant being included in the first contiguous agent, which contains the feeding elements of the right branch on the network, along with $U_4$ and $B_2$. Both Agents 2 and 3 are noncontiguous, with Agent 2 containing the remaining feeding pipe in the left branch and $U_3$. Agent 3 has the remaining two users, one from each network branch, and $B_1$. 

Additionally, the results are compared to a baseline partition, pictured in \cref{fig:part_base}. This baseline is derived by finding the modularity-maximizing partition of the component connection graph using unweighted edges \cite{jogwarCommunitybasedSynthesisDistributed2017}. Since all edges leaving the heating plant carry equal weight, the method provides no systematic basis for assigning plant control, and it is manually assigned to Agent 1. Because the baseline partitioning method works to minimize the connections between cut components in an unweighted graph, it does not change as the system parameters are modified. This method results in a partition with three contiguous subsystems, where the plant is part of the left branch with $U_2$ and $U_3$. 
\subsection{Case Descriptions} The cases considered are as follows. The first two cases change the supply temperature $T_0$ by $- 5\degree C$ (b), and $+ 5\degree C$ (c). The following two cases look at changing the operating season to the first week of November (d) and December (e). The average ambient temperature for each month was 2.4\degree C and 2.2\degree C, respectively. The buildings' demand profiles for these cases are shown in \cref{fig:demand_seasonal}. The next set of cases changes the allowable temperature deviation in the buildings, $\Delta T_B$ by $- 0.2\degree C$ (f), and $+0.2\degree C$ (g). Additional cases consider increasing and decreasing the heat transfer coefficient, $h$, of the pipes by $-0.2\ W/m^2K$ (h) and $+ 0.2\ W/m^2K$ (i) and the feeding and return pipe diameter, $D$, by $- 0.5\ m$ (j) and $- 0.5\ m$ (k). Finally, the last two cases consider changing the building type from residential to commercial for a single user in the network. The demand profile and heat capacity for this building are changed to the ones labeled "new" in \cref{fig:demand} and \cref{tbl:network}, respectively. The first of these cases considers changing $U_1$ (l) while the second considers changing $U_2$ (m). 
\begin{figure}[p]
    \centering
    \scriptsize
    \captionsetup[subfigure]{font=scriptsize,skip=1pt}

    \begin{subfigure}[t]{0.44\linewidth}
        \centering
        \includegraphics[width=\linewidth]{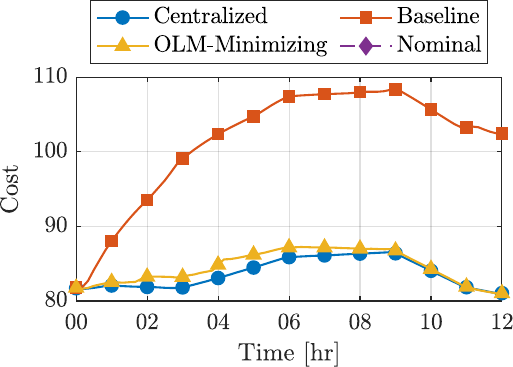}
        \caption{Nominal.}
        \label{fig:cost_nom}
    \end{subfigure}
    \hfill
    \begin{subfigure}[t]{0.44\linewidth}
        \centering
        \includegraphics[width=\linewidth]{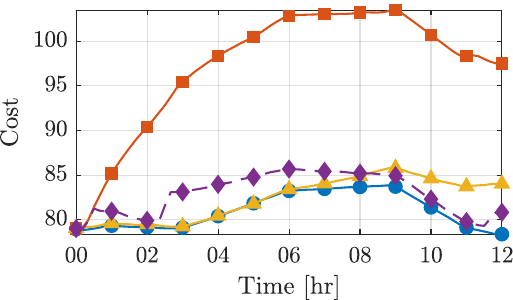}
        \caption{$T_0 = 75^\circ$C.}
        \label{fig:cost_75}
    \end{subfigure}

    \vspace{0.15em}

    \begin{subfigure}[t]{0.44\linewidth}
        \centering
        \includegraphics[width=\linewidth]{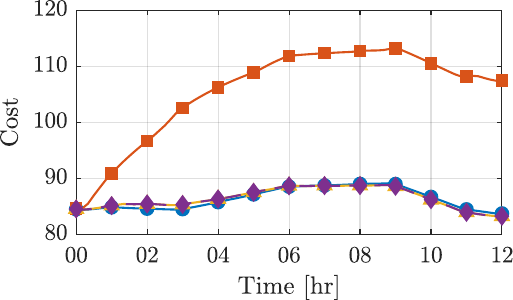}
        \caption{$T_0 = 85^\circ$C.}
        \label{fig:cost_85}
    \end{subfigure}
    \hfill
    \begin{subfigure}[t]{0.44\linewidth}
        \centering
        \includegraphics[width=\linewidth]{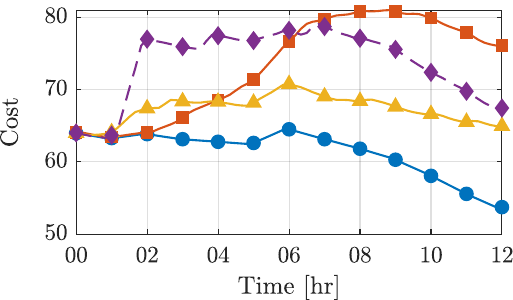}
        \caption{November.}
        \label{fig:cost_nov}
    \end{subfigure}

    \vspace{0.15em}

    \begin{subfigure}[t]{0.44\linewidth}
        \centering
        \includegraphics[width=\linewidth]{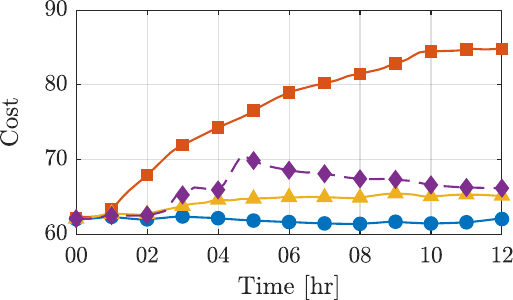}
        \caption{December.}
        \label{fig:cost_dec}
    \end{subfigure}
    \hfill
    \begin{subfigure}[t]{0.44\linewidth}
        \centering
        \includegraphics[width=\linewidth]{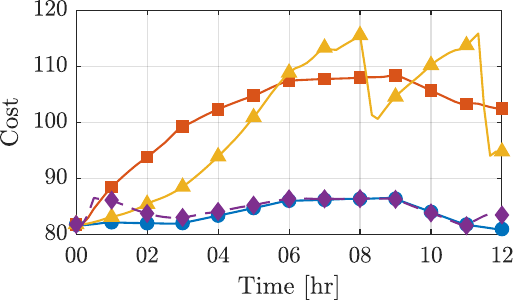}
        \caption{$\Delta T_B = 1.8^\circ$C.}
        \label{fig:cost_cbl}
    \end{subfigure}

    \vspace{0.15em}

    \begin{subfigure}[t]{0.44\linewidth}
        \centering
        \includegraphics[width=\linewidth]{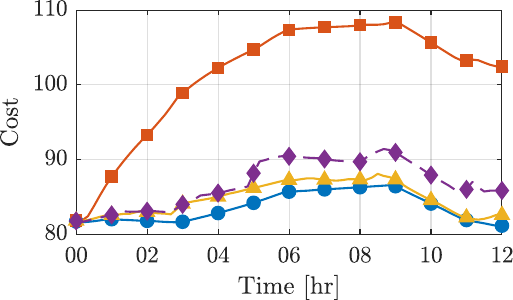}
        \caption{$\Delta T_B = 2.2^\circ$C.}
        \label{fig:cost_cbh}
    \end{subfigure}
    \hfill
    \begin{subfigure}[t]{0.44\linewidth}
        \centering
        \includegraphics[width=\linewidth]{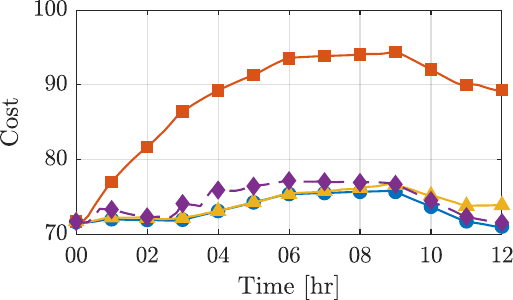}
        \caption{$h = 1.3\ \mathrm{W/m^2K}$.}
        \label{fig:cost_hl}
    \end{subfigure}

    \caption{Total cost comparison for the nominal, temperature, seasonal, building flexibility, and low heat transfer coefficient cases.}
    \label{fig:cost_all_1}
\end{figure}

\begin{figure}[p]
    \ContinuedFloat
    \centering
    \scriptsize
    \captionsetup[subfigure]{font=scriptsize,skip=1pt}

    \begin{subfigure}[t]{0.44\linewidth}
        \centering
        \includegraphics[width=\linewidth]{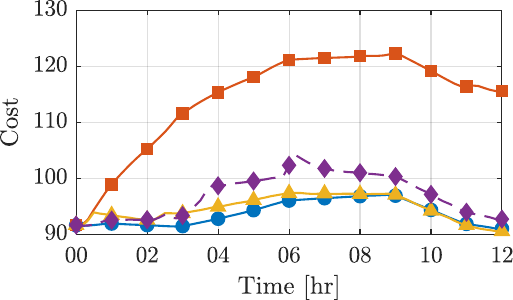}
        \caption{$h = 1.7\ \mathrm{W/m^2K}$.}
        \label{fig:cost_hh}
    \end{subfigure}
    \hfill
    \begin{subfigure}[t]{0.44\linewidth}
        \centering
        \includegraphics[width=\linewidth]{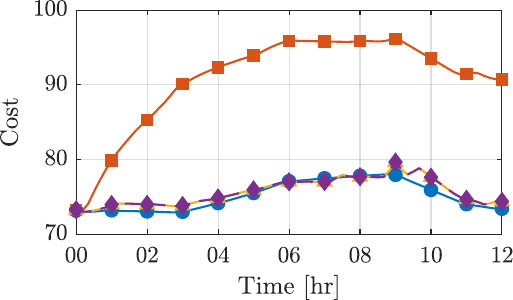}
        \caption{$D = 0.35\ \mathrm{m}$.}
        \label{fig:cost_Dl}
    \end{subfigure}

    \vspace{0.15em}

    \begin{subfigure}[t]{0.44\linewidth}
        \centering
        \includegraphics[width=\linewidth]{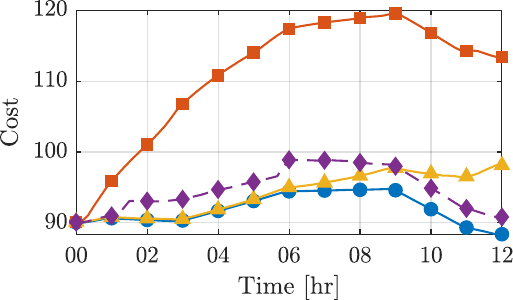}
        \caption{$D = 0.40\ \mathrm{m}$.}
        \label{fig:cost_Dh}
    \end{subfigure}
    \hfill
    \begin{subfigure}[t]{0.44\linewidth}
        \centering
        \includegraphics[width=\linewidth]{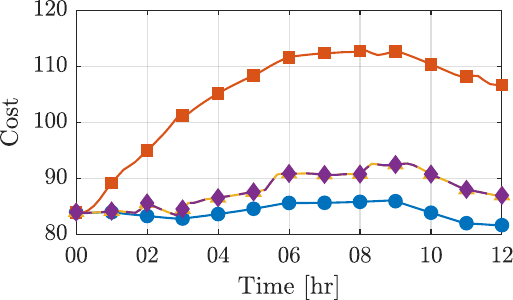}
        \caption{Commercial $U_1$.}
        \label{fig:cost_comm1}
    \end{subfigure}

    \vspace{0.15em}

    \begin{subfigure}[t]{0.44\linewidth}
        \centering
        \includegraphics[width=\linewidth]{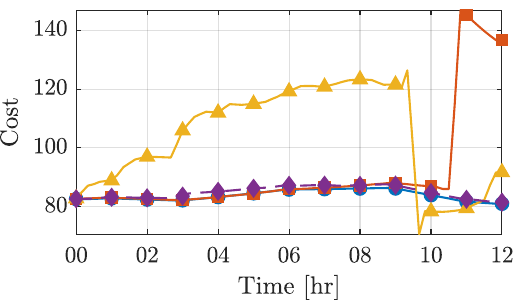}
        \caption{Commercial $U_2$.}
        \label{fig:cost_comm2}
    \end{subfigure}

    \caption{Continued total cost comparison for pipe diameter and commercial building cases.}
    \label{fig:cost_all_2}
\end{figure}
\begin{table}
\caption{Percent increase in cost from centralized control.}
\label{tbl:cost}
\centering
\begin{tabular}{c l c c c}
\toprule
Case & Variable & OLM & Nominal& Baseline\\
\midrule
a & Nominal & 1.0 & - & 21.3 \\
b & $T_0 = 75\degree C$ & 1.6 & 2.4 & 20.1\\
c & $T_0 = 85\degree C$ & 0.1 & - & 22.3\\
d & November & 9.6 & 20.1 & 19.2\\
e & December & 4.2 & 7.2 & 24.0\\
f & $\Delta T_\text{B} = 1.8\degree C$ & 19.1 & 1.1 & 21.2\\
g & $\Delta T_\text{B} = 2.2\degree C$ & 1.4 & 3.8 & 21.3\\
h & $h = 1.3\ W/m^2 K$ & 0.9 & 1.7 & 20.5\\
i & $h = 1.7\ W/m^2 K$ & 1.0 & 3.5 & 22.0\\
j & $D=0.35\ m$ & 0.7 & - & 20.9\\
k & $D=0.45\ m$ & 2.3 & 3.1 & 20.7\\
l & New $U_1$ & 4.7 & - & 24.8\\
m & New $U_2$ & 23.8 & 1.3 & 9.0\\
\bottomrule
\end{tabular}
\end{table}
\subsection{Results}
The results of the sensitivity analysis are presented in \cref{fig:cost_all_1,fig:cost_all_2,fig:part_all_1,fig:part_all_2,tbl:cost}.
\Cref{fig:cost_all_1}, \cref{fig:cost_all_2} and \cref{tbl:cost} present the cost over time and the overall percent increase in cost for both the nominal and case-specific OLM-minimizing-partition. The OLM-minimizing partitions found by the heuristic algorithm for each case are presented in \cref{fig:part_all_1} and \cref{fig:part_all_2}.
Additionally, \cref{apx:fig} presents the used flexibilities and plant mass flow rates for these controllers. The sensitivity analysis on these cases reveals two key results: 
\begin{enumerate}
    \item A well-designed partition demonstrates strong robustness to a wide range of parameter variations, with meaningful performance degradation observed only under significant seasonal demand shifts.
    \item The effectiveness of the OLM evaluation is sensitive to the selection of cost function weights and initial conditions, with 5 of 12 case-specific partitions underperforming the nominal despite being identified as OLM-minimizing. 
\end{enumerate}
\begin{figure}[p]
    \centering
    \scriptsize
    \captionsetup[subfigure]{font=scriptsize,skip=1pt}

    \begin{subfigure}[t]{0.44\linewidth}
        \centering
        \includegraphics[width=\linewidth, trim={0 31 0 4}, clip]{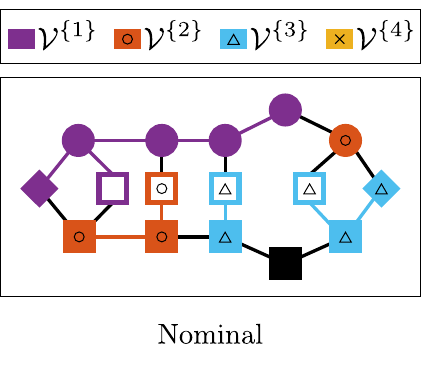}
        \caption{Nominal.}
        \label{fig:part_nom}
    \end{subfigure}
    \hfill
    \begin{subfigure}[t]{0.44\linewidth}
        \centering
        \includegraphics[width=\linewidth, trim={0 31 0 4}, clip]{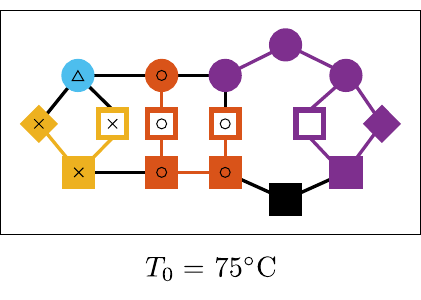}
        \caption{$T_0 = 75^\circ$C.}
        \label{fig:part_75}
    \end{subfigure}

    \vspace{0.15em}

    \begin{subfigure}[t]{0.44\linewidth}
        \centering
        \includegraphics[width=\linewidth, trim={0 31 0 4}, clip]{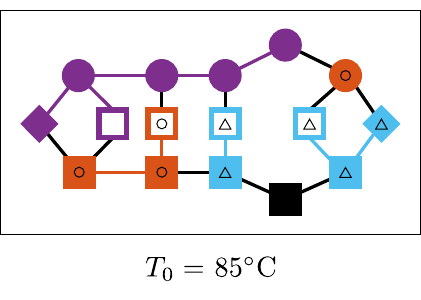}
        \caption{$T_0 = 85^\circ$C.}
        \label{fig:part_85}
    \end{subfigure}
    \hfill
    \begin{subfigure}[t]{0.44\linewidth}
        \centering
        \includegraphics[width=\linewidth, trim={0 31 0 4}, clip]{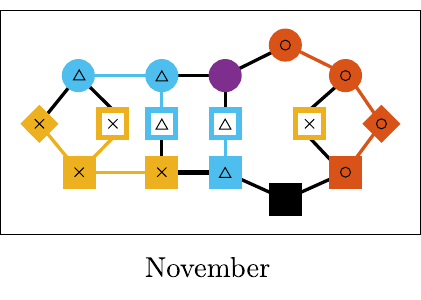}
        \caption{November.}
        \label{fig:part_nov}
    \end{subfigure}

    \vspace{0.15em}

    \begin{subfigure}[t]{0.44\linewidth}
        \centering
        \includegraphics[width=\linewidth, trim={0 31 0 4}, clip]{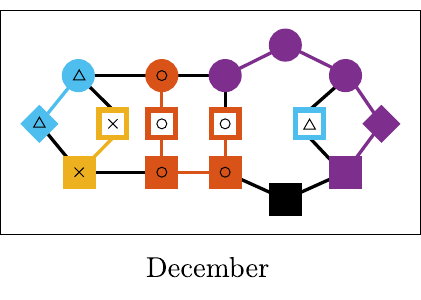}
        \caption{December.}
        \label{fig:part_dec}
    \end{subfigure}
    \hfill
    \begin{subfigure}[t]{0.44\linewidth}
        \centering
        \includegraphics[width=\linewidth, trim={0 31 0 4}, clip]{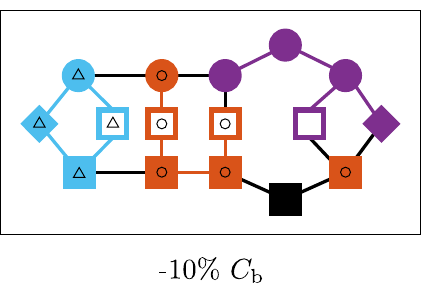}
        \caption{$\Delta T_B = 1.8^\circ$C.}
        \label{fig:part_cbl}
    \end{subfigure}

    \vspace{0.15em}

    \begin{subfigure}[t]{0.44\linewidth}
        \centering
        \includegraphics[width=\linewidth, trim={0 31 0 4}, clip]{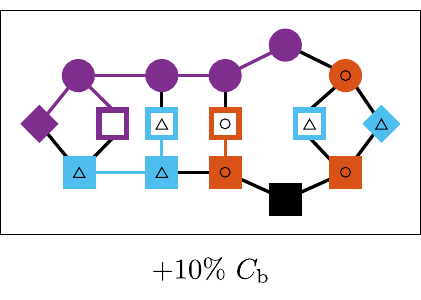}
        \caption{$\Delta T_B = 2.2^\circ$C.}
        \label{fig:part_cbh}
    \end{subfigure}
    \hfill
    \begin{subfigure}[t]{0.44\linewidth}
        \centering
        \includegraphics[width=\linewidth, trim={0 31 0 4}, clip]{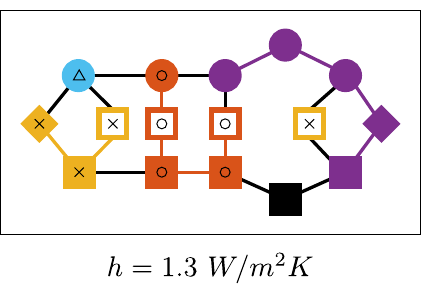}
        \caption{$h = 1.3~\mathrm{W/m^2K}$.}
        \label{fig:part_hl}
    \end{subfigure}

    \caption{OLM-minimizing partition for the nominal, temperature, seasonal, building flexibility, and low heat transfer coefficient cases.}
    \label{fig:part_all_1}
\end{figure}

\begin{figure}[p]
    \ContinuedFloat
    \centering
    \scriptsize
    \captionsetup[subfigure]{font=scriptsize,skip=1pt}

    \begin{subfigure}[t]{0.44\linewidth}
        \centering
        \includegraphics[width=\linewidth, trim={0 31 0 4}, clip]{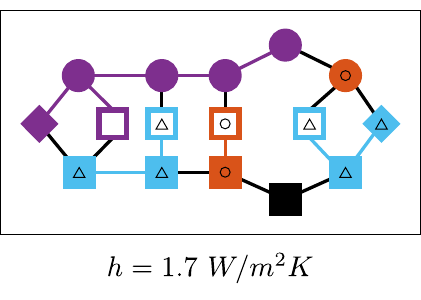}
        \caption{$h = 1.7~\mathrm{W/m^2K}$.}
        \label{fig:part_hh}
    \end{subfigure}
    \hfill
    \begin{subfigure}[t]{0.44\linewidth}
        \centering
        \includegraphics[width=\linewidth, trim={0 31 0 4}, clip]{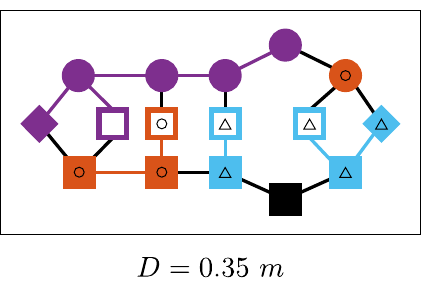}
        \caption{$D = 0.35~\mathrm{m}$.}
        \label{fig:part_Dl}
    \end{subfigure}

    \vspace{0.15em}

    \begin{subfigure}[t]{0.44\linewidth}
        \centering
        \includegraphics[width=\linewidth, trim={0 31 0 4}, clip]{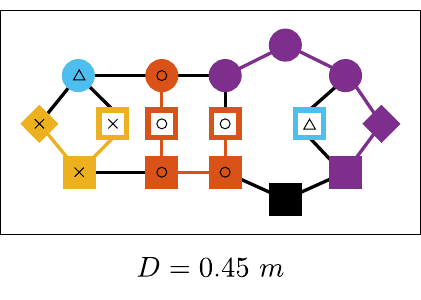}
        \caption{$D = 0.40~\mathrm{m}$.}
        \label{fig:part_Dh}
    \end{subfigure}
    \hfill
    \begin{subfigure}[t]{0.44\linewidth}
        \centering
        \includegraphics[width=\linewidth, trim={0 31 0 4}, clip]{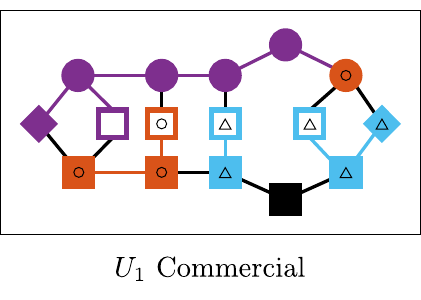}
        \caption{Commercial $U_1$.}
        \label{fig:part_comm1}
    \end{subfigure}

    \vspace{0.15em}

    \begin{subfigure}[t]{0.44\linewidth}
        \centering
        \includegraphics[width=\linewidth, trim={0 31 0 4}, clip]{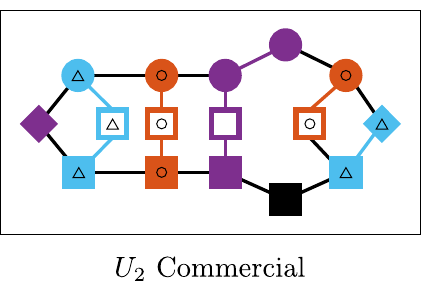}
        \caption{Commercial $U_2$.}
        \label{fig:part_comm2}
    \end{subfigure}

    \caption{Continued OLM-minimizing partition for high heat transfer coefficient, pipe diameter, and commercial building cases.}
    \label{fig:part_all_2}
\end{figure}

\paragraph{Result 1} 
From \cref{fig:cost_all_1,fig:cost_all_2,tbl:cost}, it is evident that the nominal partition performed well in 11 of the 12 cases considered. The average percent cost increase from the centrally optimal solution in these 11 cases was only 2.8\%. Moreover, in three cases: increased supply temperature (c), decreased diameter (j), and commercial $U_1$ (l), the OLM evaluation confirms the nominal partition as optimal even under the perturbed conditions, demonstrating that these parameter variations do not require any structural changes to the partition. In two additional cases, increased building temperature deviation (g) and worse insulation (i), the OLM identifies only minor modifications to the nominal partition, with an identical agent one, and two additional noncontiguous agents incorporating small component swaps. This indicates that moderate changes in building flexibility and pipe insulation can be accommodated with minimal changes to the system partitioning. This demonstrates that a partition designed for a single representative operating point can maintain near-optimal performance across a substantial range of system conditions.\par
The only case where the nominal partition experiences meaningful performance degradation is in November (\cref{fig:cost_nov}). In November, the nominal partition performs worse than the baseline method, indicating that seasonal repartitioning would improve control performance. The building heat demands in November have changed by an average of 61\% relative to the nominal January conditions, necessitating this controller redesign. The OLM-minimizing partitions for both November and December show significant structural changes from the nominal, including reorganization into four agents rather than three, further supporting the need for seasonal adaptation. However, the absolute losses and plant flow in these warmer months are lower than in other cases, so while repartitioning is beneficial, the absolute performance impact remains modest. \par

\paragraph{Result 2}
In five cases, the OLM-minimizing partition underperforms the nominal. In two cases: decrease allowable temperature deviation and a commercial $U_2$, the OLM-minimizing partition performs consistently worse throughout the simulation. In three cases: decreased supply temperature, decreased insulation, and increased diameter, the OLM partitions initially show improved performance, but by the end of the 12-hour simulation, their costs exceed those of the nominal controller. To confirm the heuristic was not at fault, the OLM was calculated for the nominal partition in these cases, verifying that it has a higher $c_{\text{olm}}$ than the found partitions for these cases.\par
In all of these poor-performing cases, the performance degradation follows a consistent pattern. The found OLM-minimizing partition drives building flexibility to saturation early in the simulation horizon, but lacks the communication structure necessary to coordinate the heat delivery between buildings to enable oscillations around these limits. In contrast, the centralized and nominal partitions maintain this coordination, allowing charge-sustaining behavior that continues to reduce losses after the buildings have reached this saturation. This lack of long-term coordination ability comes from the structural changes in the OLM-minimizing partition. In all of these cases except m, these partitions have fewer noncontiguous agents and exhibit more localized grouping, where agents are organized vertically through feeding-user-return paths rather than horizontally across feeding lines. This resembles the baseline partitioning approach and reduces coordination among the agents. While the partition in case m has three noncontiguous agents, it still exhibits these vertically oriented agents.\par
\begin{figure}
    \centering
    \begin{subfigure}[t]{.49\linewidth}
        \includegraphics[width=\linewidth, trim={0 31 0 4}, clip]{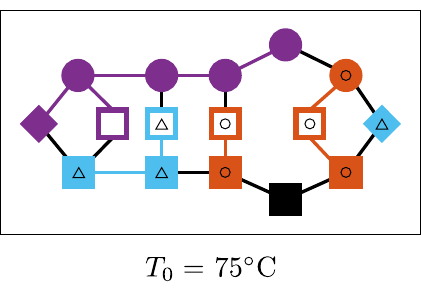}
        \caption{OLM-minimizing partition.}
        \label{fig:part_weight}
    \end{subfigure}
    \begin{subfigure}[t]{.49\linewidth}
        \includegraphics[width=\linewidth]{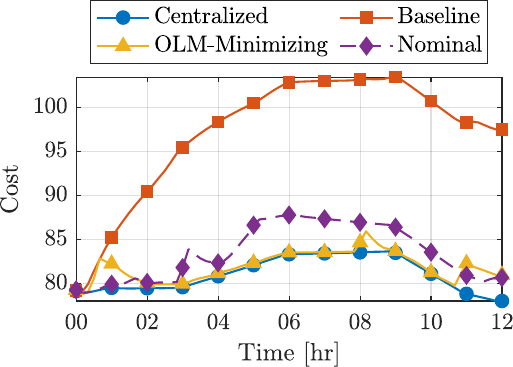}
        \caption{Total cost.}
        \label{fig:cost_weight}
    \end{subfigure}\\
    \vspace{10pt}%
    \begin{subfigure}[t]{.95\linewidth}
        \includegraphics[width=\linewidth]{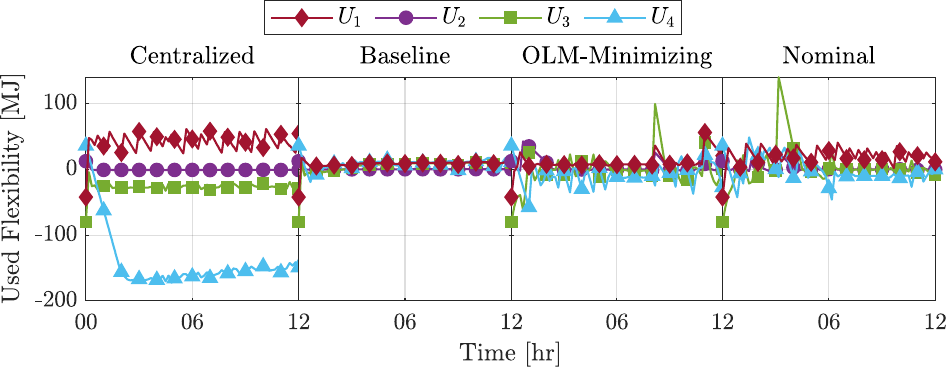}
        \caption{Used flexibility.}
        \label{fig:flex_weight}
    \end{subfigure}
    \caption{Results from re-tuning the weights for  $T_0=75\degree C$.}
    \label{fig:newWeight}
\end{figure}

The underlying cause of this coordination failure is that the parameter changes alter the relative magnitude of heat losses to flexibility costs in the system. These deviations affect the ability of the cost function in \cref{eq:i_cost} to effectively capture the long-term disadvantage of using building flexibility to reduce heat losses. In the single-step OLM calculation, if using flexibility incurs little immediate penalty, the optimization does not identify the communication structure necessary for long-term coordination of heat demands. Without retuning the cost function weights in \cref{eq:i_cost} or adjusting the initial conditions for the building states of energy, the single-step OLM evaluation optimizes for the initial operating point but fails to identify partitions that maintain coordination over the full horizon. The nominal partition, for which cost weights and initial conditions were explicitly tuned to the January operating conditions, maintains effective coordination even under parameter variations. 
To demonstrate this effect, the partition optimization was repeated for case b, where $T_0 = 75 \degree C$, with the weight on the used flexibility term $w_C$ increased from 5 to 10. The results of this partitioning are shown in \cref{fig:newWeight}. The flexibility results in \cref{fig:flex_weight} show that when using the optimal partition for the new term weighting, the flexibility no longer saturates, indicating that the observed underperformance of the OLM-minimizing partition is not due to a structural limitation of the partitioning framework, but rather to the parametrization of the cost function used in the single-step evaluation. This is reflected in the simulation-long improved costs shown in \cref{fig:cost_weight}. The partition, presented in \cref{fig:part_weight}, found for the new cost weight is also more similar to the other effective partitions, exhibiting the horizontal structure that enables effective communication between subsystems. By increasing the relative weight on flexibility usage, the resulting partition avoids early saturation and recovers coordinated behavior over the full simulation horizon. This confirms that the sensitivity observed in these cases is driven by the balance between flexibility and heat-loss penalties and that proper tuning yields partitions that maintain robustness across operational variations, as demonstrated in Result 1.

\section{Conclusion}
\label{sec:conclusion}
This paper presented a sensitivity analysis of the factors affecting the optimal partitioning of a DHN for distributed control. Using a communication-based distributed model predictive control framework and a learning-enhanced branch and bound heuristic, the effects of twelve parameter variations were examined relative to a nominal operating case. These case studies demonstrated that a well-designed partition is highly robust to a wide range of parameter perturbations. The only case with meaningful performance degradation was the November operating season, indicating that seasonal repartitioning is warranted when demand profiles change substantially. However, the OLM metric was found to be sensitive to cost function weights and initial conditions. Because the cost weights and initial conditions were explicitly tuned for the nominal case study, the OLM metric identified a partition that enabled effective coordination between agents across nearly all tested conditions. These findings indicate that proper tuning of the controller and initial conditions for the design case study is essential for reliable partition selection and that the nominal partition, when carefully designed, offers strong robustness as a fixed control architecture.\par
Future work will explore two main directions. First is addressing the super-exponential growth of the partitioning solution space. The consistent structure of well-performing partitions identified in this sensitivity analysis, having horizontal groupings along feeding lines rather than vertical feeding, user, and return groups, suggests a promising direction for scaling the partitioning framework to larger networks. By using this structural insight to generate an initial partition, the $\mathcal{NP}$-hard search can be recast as a local refinement problem, where OLM is used to evaluate element-wise changes to this initial partition. This would reduce the effective search space from exponential in network size to polynomial in the number of refinement steps, making the framework tractable for larger DHNs. Additionally, by incorporating physical considerations, such as feasible communication distances between system elements and the location of booster pumps, which decouple pressure constraints, this performance-based partitioning can scale to full-size DHNs. Communication distance constraints prune the combinatorial search space to realizable configurations, while booster pump locations provide natural hydraulic decoupling points that allow each subsystem to satisfy pressure feasibility constraints independently. The other will focus on the main limitation identified in this paper: the need to re-tune the cost function weights as the system parameters change. This work will investigate the problem formulation needed to make the OLM evaluation sufficiently expressive across the full range of DHN operating conditions. It will focus on defining case studies that stress the system effectively, identifying the set of design scenarios that yield well-performing partitions for all anticipated operating conditions in the fewest evaluations possible.

\bibliographystyle{unsrt}
\bibliography{sources}

@misc{blizardOptimalityLossMinimization2025b,
  title = {Optimality {{Loss Minimization}} in {{Distributed Control}} with {{Application}} to {{District Heating}}},
  author = {Blizard, Audrey and Stockar, Stephanie},
  year = {2025},
  month = jul,
  number = {arXiv:2507.02144},
  eprint = {2507.02144},
  primaryclass = {eess},
  publisher = {arXiv},
  doi = {10.48550/arXiv.2507.02144},
  urldate = {2025-08-08},
  archiveprefix = {arXiv}
}

@article{seatonIntrinsicFragilityPrice2023,
  title = {On the {{Intrinsic Fragility}} of the {{Price}} of {{Anarchy}}},
  author = {Seaton, Joshua H. and Brown, Philip N.},
  year = {2023},
  journal = {IEEE Control Systems Letters},
  volume = {7},
  pages = {3573--3578},
  issn = {2475-1456},
  doi = {10.1109/LCSYS.2023.3335315},
  urldate = {2024-05-13}
}

@article{shiNormalizedCutsImage2000,
  title = {Normalized Cuts and Image Segmentation},
  author = {Shi, Jianbo and Malik, J.},
  year = {2000},
  month = aug,
  journal = {IEEE Transactions on Pattern Analysis and Machine Intelligence},
  volume = {22},
  number = {8},
  pages = {888--905},
  issn = {1939-3539},
  doi = {10.1109/34.868688}
}

@inproceedings{blizardAcceleratingDistributedControl2025,
  title = {Accelerating {{Distributed Control Design}} for {{District Heating Networks}} via {{Learning}} of {{Critical Communication Links}}},
  booktitle = {{{IFAC-PapersOnLine}}},
  author = {Blizard, Audrey and Stockar, Stephanie},
  year = 2025,
  month = oct,
  series = {5th {{Conference}} on {{Modeling}}, {{Estimation}} and {{Control MECC}} 2025},
  volume = {59},
  pages = {419--424},
  publisher = {IFAC},
  address = {Pittsburgh, Pennsylvania},
  issn = {2405-8963},
  doi = {10.1016/j.ifacol.2025.12.273}
}

@book{jamesIntroductionStatisticalLearning2013,
  title = {An Introduction to Statistical Learning: With Applications in {{R}}},
  shorttitle = {An Introduction to Statistical Learning},
  editor = {James, Gareth and Witten, Daniela and Hastie, Trevor and Tibshirani, Robert},
  year = 2013,
  series = {Springer Texts in Statistics},
  number = {103},
  publisher = {Springer},
  address = {New York},
  isbn = {978-1-4614-7137-0},
  lccn = {QA276 .I585 2013}
}

@Article{Andersson2019,
  author = {Joel A E Andersson and Joris Gillis and Greg Horn
            and James B Rawlings and Moritz Diehl},
  title = {{CasADi} -- {A} software framework for nonlinear optimization
           and optimal control},
  journal = {Mathematical Programming Computation},
  volume = {11},
  number = {1},
  pages = {1--36},
  year = {2019},
  publisher = {Springer},
  doi = {10.1007/s12532-018-0139-4}
}

@article{wachterImplementationInteriorpointFilter2006,
  title = {On the Implementation of an Interior-Point Filter Line-Search Algorithm for Large-Scale Nonlinear Programming},
  author = {W{\"a}chter, Andreas and Biegler, Lorenz T.},
  year = {2006},
  month = mar,
  journal = {Mathematical Programming},
  volume = {106},
  number = {1},
  pages = {25--57},
  issn = {1436-4646},
  doi = {10.1007/s10107-004-0559-y},
  urldate = {2023-11-27},
  langid = {english}
}

@article{jogwarCommunitybasedSynthesisDistributed2017,
  title = {Community-Based Synthesis of Distributed Control Architectures for Integrated Process Networks},
  author = {Jogwar, Sujit Suresh and Daoutidis, Prodromos},
  year = {2017},
  month = nov,
  journal = {Chemical Engineering Science},
  volume = {172},
  pages = {434--443},
  issn = {0009-2509},
  doi = {10.1016/j.ces.2017.06.043},
  urldate = {2022-07-01},
  langid = {english}
}

@article{ebrahimiAdaptiveDistributedArchitecture2024,
  title = {An Adaptive Distributed Architecture for Multi-Agent State Estimation and Control of Complex Process Systems},
  author = {Ebrahimi, AmirMohammad and Pourkargar, Davood B.},
  year = 2024,
  month = oct,
  journal = {Chemical Engineering Research and Design},
  volume = {210},
  pages = {594--604},
  issn = {0263-8762},
  doi = {10.1016/j.cherd.2024.09.014}
}

@article{wackMultiperiodTopologyDesign2024,
  title = {A Multi-Period Topology and Design Optimization Approach for District Heating Networks},
  author = {Wack, Yannick and Sollich, Martin and Salenbien, Robbe and Diriken, Jan and Baelmans, Martine and Blommaert, Maarten},
  year = 2024,
  month = aug,
  journal = {Applied Energy},
  volume = {367},
  pages = {123380},
  issn = {0306-2619},
  doi = {10.1016/j.apenergy.2024.123380}
}

@article{murosGameTheoreticalRandomized2018,
  title = {A {{Game Theoretical Randomized Method}} for {{Large-Scale Systems Partitioning}}},
  author = {Muros, Francisco Javier and Maestre, Jos{\'e} Mar{\'i}a and {Ocampo-Martinez}, Carlos and Algaba, Encarnaci{\'o}N and Camacho, Eduardo F.},
  year = 2018,
  journal = {IEEE Access},
  volume = {6},
  pages = {42245--42263},
  issn = {2169-3536},
  doi = {10.1109/ACCESS.2018.2854783}
}

@article{pourkargarComprehensiveStudyDecomposition2018,
  title = {Comprehensive Study of Decomposition Effects on Distributed Output Tracking of an Integrated Process over a Wide Operating Range},
  author = {Pourkargar, Davood Babaei and Almansoori, Ali and Daoutidis, Prodromos},
  year = 2018,
  month = jun,
  journal = {Chemical Engineering Research and Design},
  volume = {134},
  pages = {553--563},
  issn = {0263-8762},
  doi = {10.1016/j.cherd.2018.04.045}
}

@article{tangRelativeTimeaveragedGain2018,
  title = {Relative Time-Averaged Gain Array ({{RTAGA}}) for Distributed Control-Oriented Network Decomposition},
  author = {Tang, Wentao and Babaei Pourkargar, Davood and Daoutidis, Prodromos},
  year = 2018,
  journal = {AIChE Journal},
  volume = {64},
  number = {5},
  pages = {1682--1690},
  issn = {1547-5905},
  doi = {10.1002/aic.16130}
}

@article{brandesModularityClustering2008,
  title = {On {{Modularity Clustering}}},
  author = {Brandes, Ulrik and Delling, Daniel and Gaertler, Marco and Gorke, Robert and Hoefer, Martin and Nikoloski, Zoran and Wagner, Dorothea},
  year = 2008,
  month = feb,
  journal = {IEEE Transactions on Knowledge and Data Engineering},
  volume = {20},
  number = {2},
  pages = {172--188},
  issn = {1558-2191},
  doi = {10.1109/TKDE.2007.190689}
}

@article{ocampo-martinezPartitioningApproachOriented2011,
  title = {Partitioning Approach Oriented to the Decentralised Predictive Control of Large-Scale Systems},
  author = {{Ocampo-Martinez}, C. and Bovo, S. and Puig, V.},
  year = 2011,
  month = jun,
  journal = {Journal of Process Control},
  series = {Special {{Issue}} on {{Hierarchical}} and {{Distributed Model Predictive Control}}},
  volume = {21},
  number = {5},
  pages = {775--786},
  issn = {0959-1524},
  doi = {10.1016/j.jprocont.2010.12.005}
}

@article{jogwarDistributedControlArchitecture2019,
  title = {Distributed Control Architecture Synthesis for Integrated Process Networks through Maximization of Strength of Input--Output Impact},
  author = {Jogwar, Sujit S.},
  year = 2019,
  month = nov,
  journal = {Journal of Process Control},
  volume = {83},
  pages = {77--87},
  issn = {0959-1524},
  doi = {10.1016/j.jprocont.2019.08.009}
}

@techreport{horowitz2019resstock,
  title={ResStock™},
  author={Horowitz, Scott and Pathak, Maharshi and Wilson, Eric and Merket, Noel and Christensen, Craig and Fontanini, Anthony and Horsey, Henry and Robertson, Joseph and Rager, David and Alley, John and others},
  year={2019},
  institution={National Renewable Energy Lab.(NREL), Golden, CO (United States)}
}
\appendix
\newpage \clearpage
\onecolumn

\section{Extended Results}\label{apx:fig}
This appendix presents additional figures used in the analysis of the control performance. It shows both the plant mass flow rates and the used flexibility, the two main control features of the network.
\begin{figure}[H]
    \centering 
        \begin{subfigure}[b]{.85\linewidth}
        \centering 
        \includegraphics[width=\linewidth, trim={0 0 0 0}, clip]{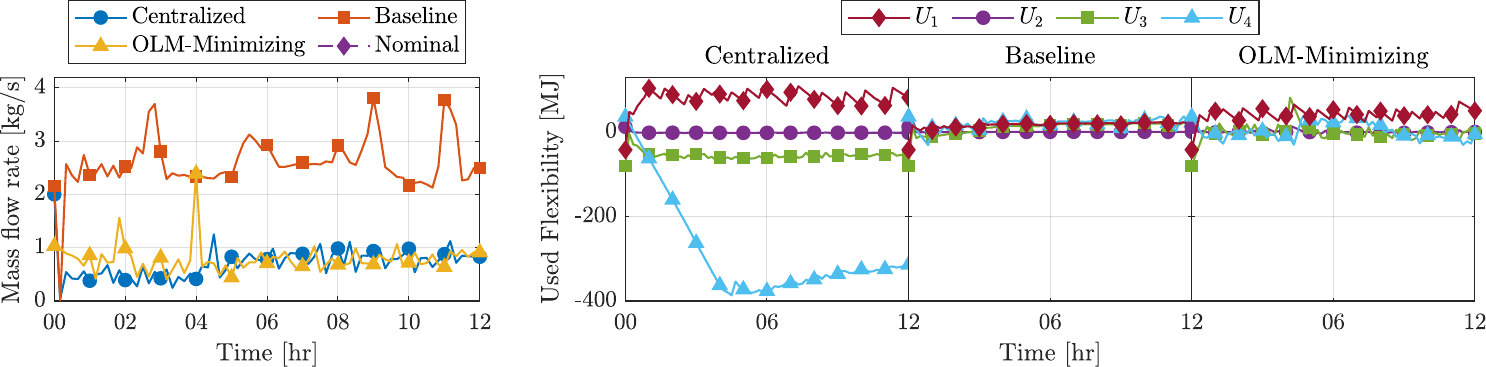}
        \caption{Nominal.}
        \label{fig:flex_nom}
    \end{subfigure}
    \begin{subfigure}[b]{.49\linewidth}
        \includegraphics[width=\linewidth, trim={0 0 0 0}, clip]{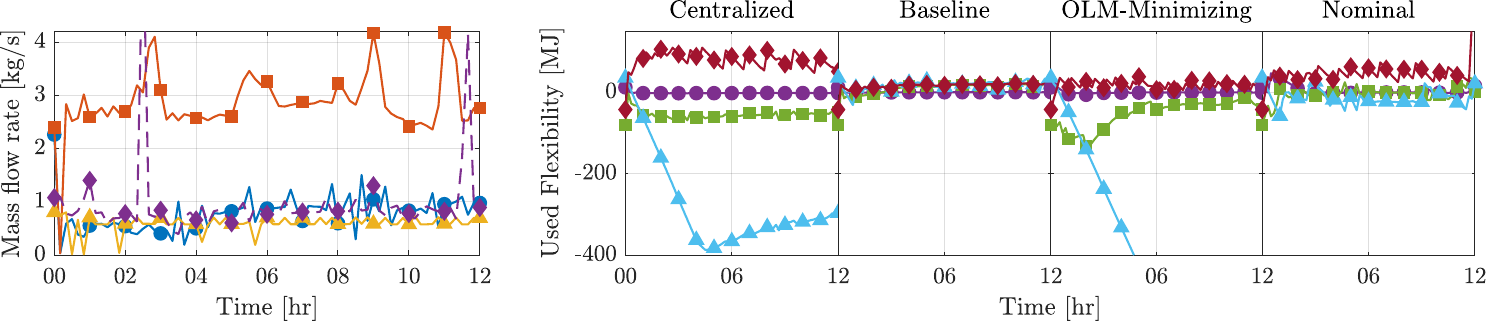}
        \caption{$T_0 = 75$\degree C.}
        \label{fig:flex_75}
    \end{subfigure}
    \hfill
    \begin{subfigure}[b]{.49\linewidth}
        \centering 
        \includegraphics[width=\linewidth, trim={0 0 0 0}, clip]{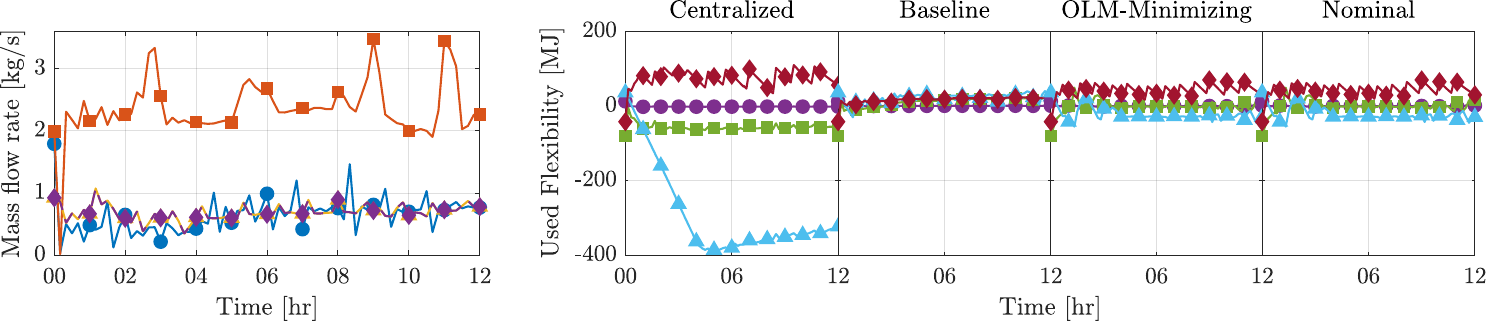}
        \caption{$T_0 = 85$\degree C.}
        \label{fig:flex_85}
    \end{subfigure}
    
    \begin{subfigure}[b]{.49\linewidth}
        \centering 
        \includegraphics[width=\linewidth, trim={0 0 0 0}, clip]{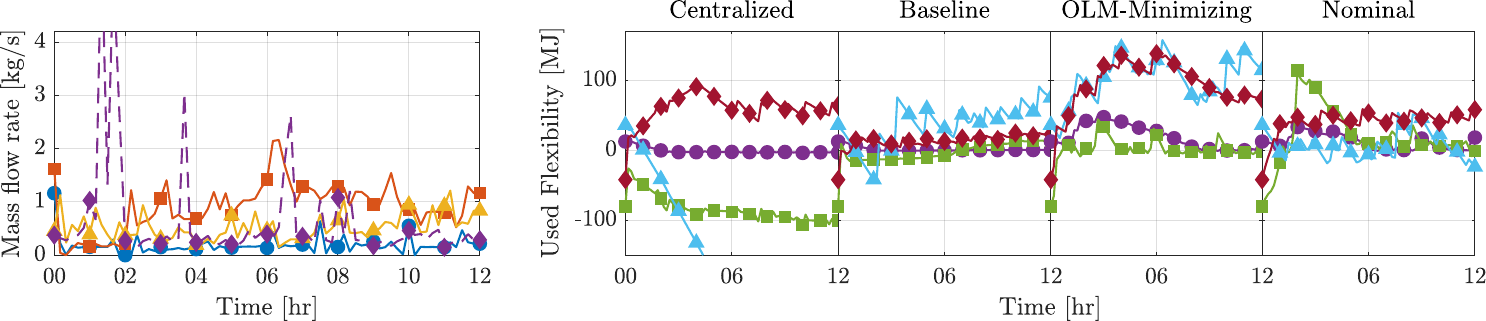}
        \caption{November.}
        \label{fig:flex_nov}
    \end{subfigure}
    \hfill
    \begin{subfigure}[b]{.49\linewidth}
        \centering 
        \includegraphics[width=\linewidth, trim={0 0 0 0}, clip]{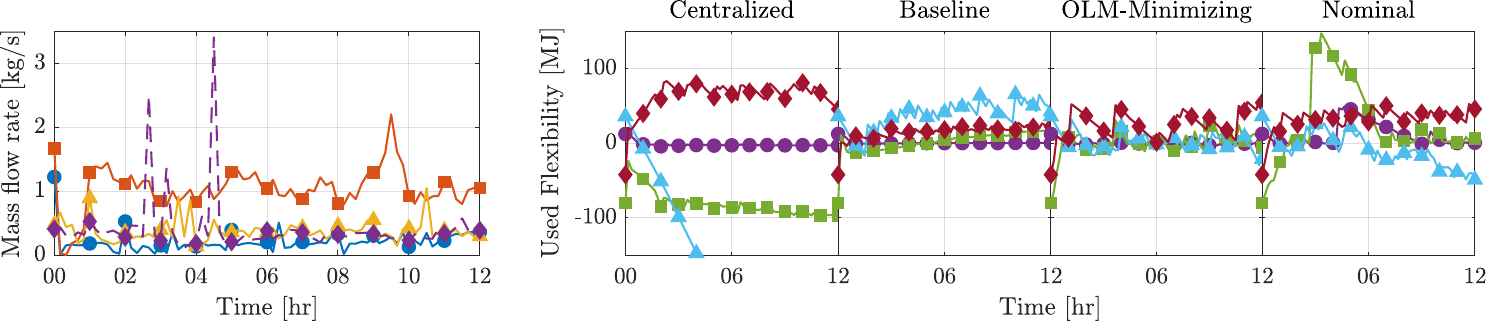}
        \caption{December.}
        \label{fig:flex_dec}
    \end{subfigure}

    \begin{subfigure}[b]{.49\linewidth}
        \centering
        \includegraphics[width=\linewidth, trim={0 0 0 0}, clip]{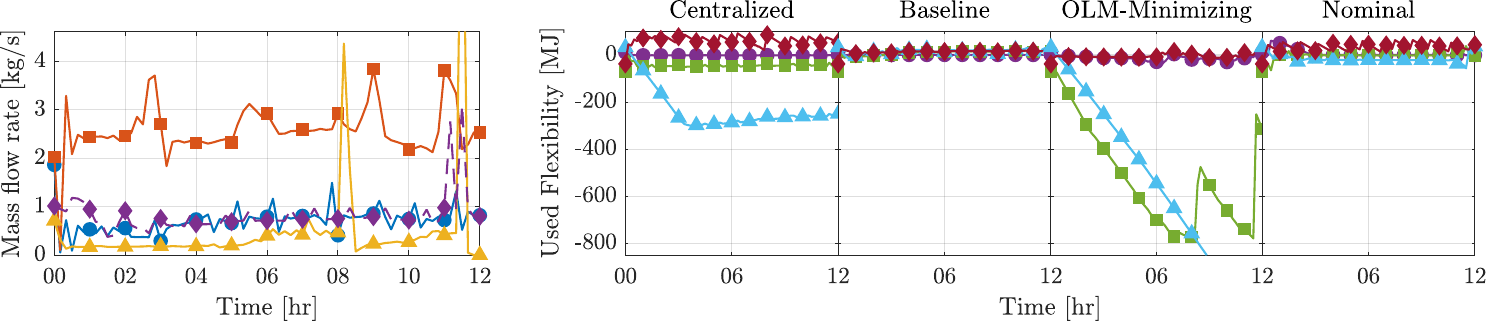}
        \caption{$\Delta T_\text{B} = 1.8\degree C$.}
        \label{fig:flex_cbl}
    \end{subfigure}
    \hfill
    \begin{subfigure}[b]{.49\linewidth}
        \centering
        \includegraphics[width=\linewidth, trim={0 0 0 0}, clip]{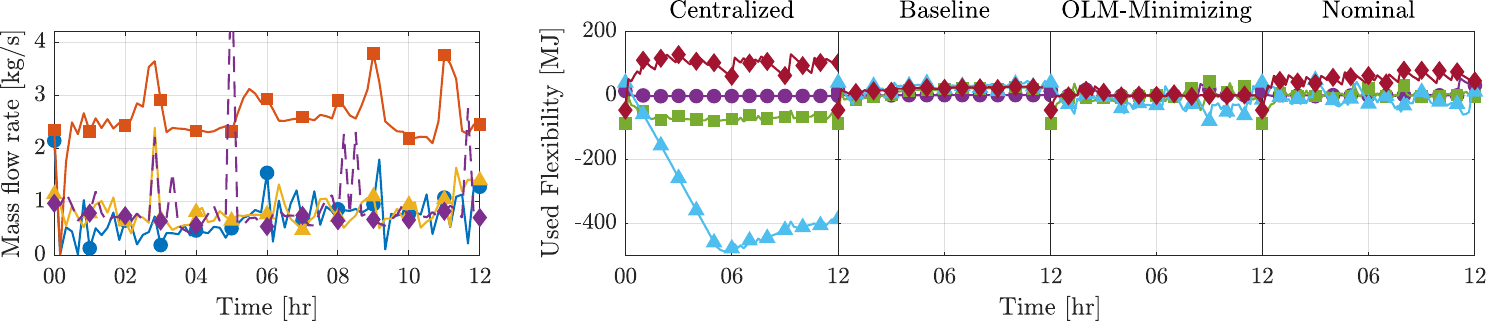}
        \caption{$\Delta T_\text{B} = 2.2\degree C$.}
        \label{fig:flex_cbh}
    \end{subfigure}

    \begin{subfigure}[b]{.49\linewidth}
        \centering
        \includegraphics[width =\linewidth, trim={0 0 0 0}, clip]{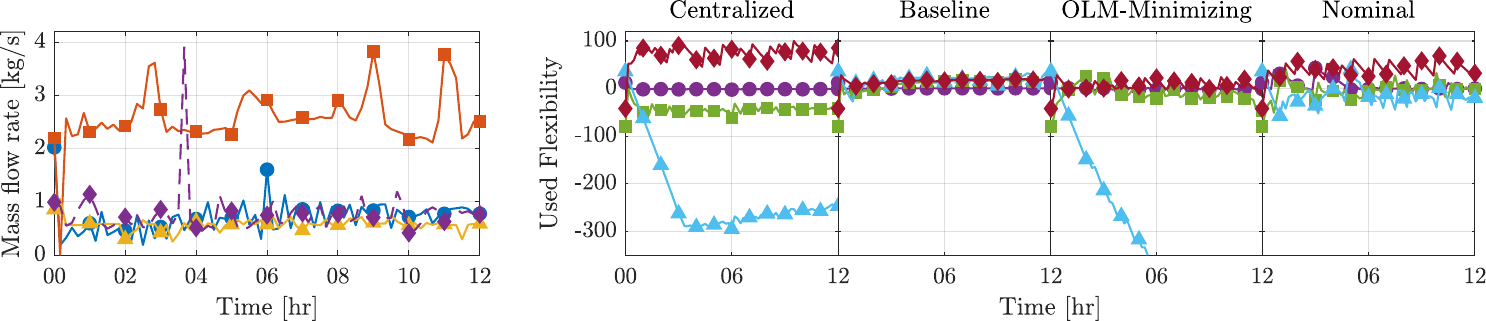}
        \caption{$h = 1.3\ W/m^2 K$.}
        \label{fig:flex_hl}
    \end{subfigure}
    \hfill
    \begin{subfigure}[b]{.49\linewidth}
        \centering
        \includegraphics[width=\linewidth, trim={0 0 0 0}, clip]{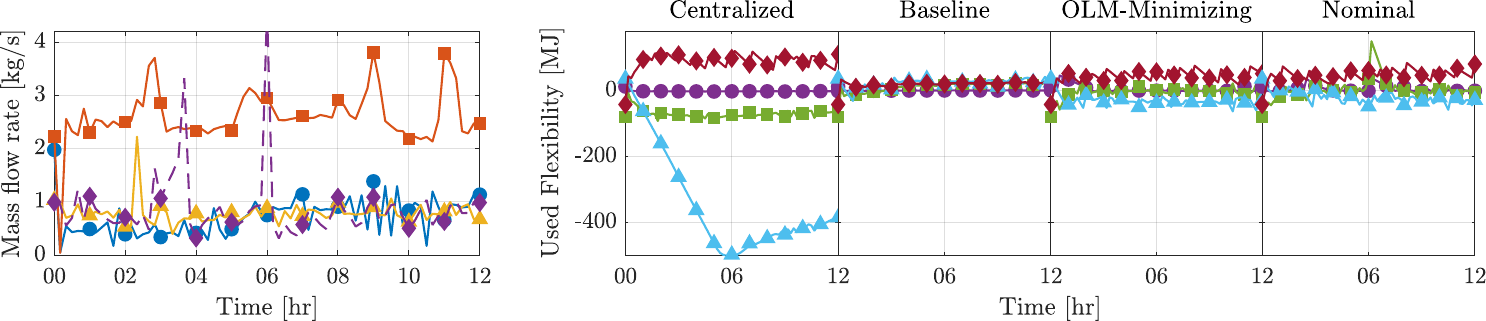}
        \caption{$h = 1.7\ W/m^2 K$.}
        \label{fig:flex_hh}
    \end{subfigure}
  
    \begin{subfigure}[b]{.49\linewidth}
        \centering
        \includegraphics[width=\linewidth, trim={0 0 0 0}, clip]{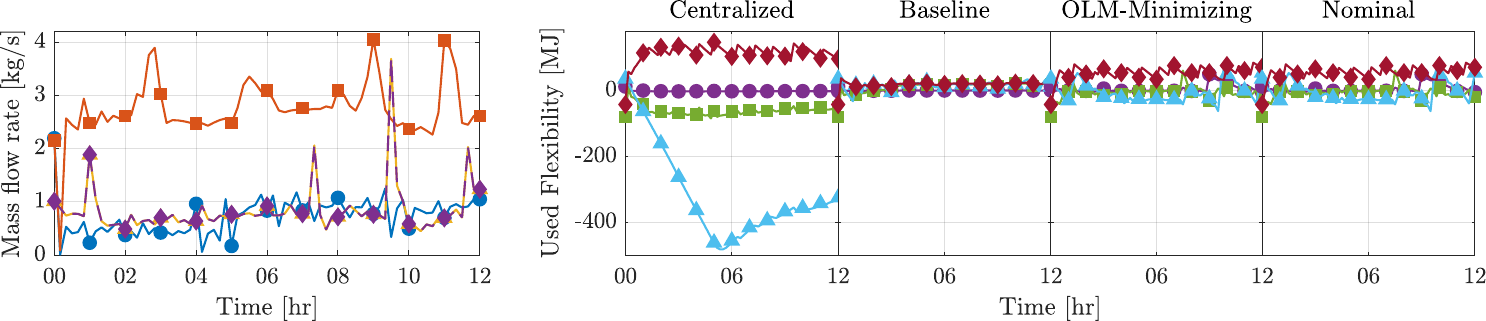}
        \caption{$D = 0.35\ m$.}
        \label{fig:flex_Dl}
    \end{subfigure}
    \hfill
    \begin{subfigure}[b]{.49\linewidth}
        \centering
        \includegraphics[width=\linewidth, trim={0 0 0 0}, clip]{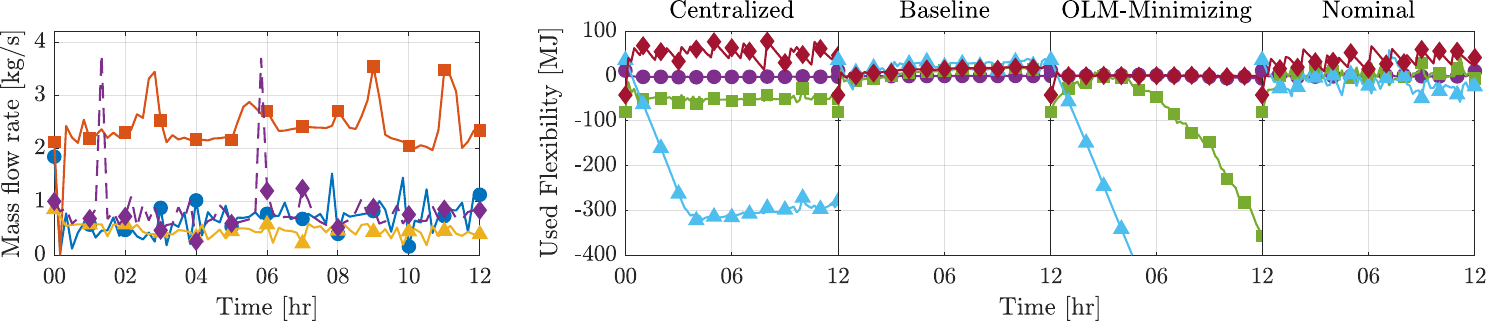}
        \caption{$D = 0.4\ m$.}
        \label{fig:flex_Dh}
    \end{subfigure}

    \begin{subfigure}[b]{.49\linewidth}
        \centering
        \includegraphics[width=\linewidth, trim={0 0 0 0}, clip]{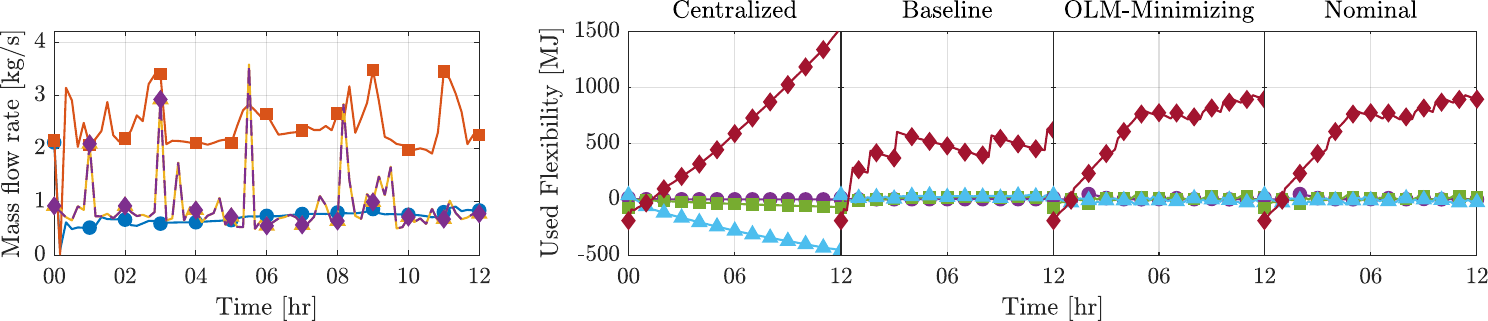}
        \caption{Commercial $U_1$.}
        \label{fig:flex_comm1}
    \end{subfigure}
    \hfill
    \begin{subfigure}[b]{.49\linewidth}
        \centering
        \includegraphics[width=\linewidth, trim={0 0 0 0}, clip]{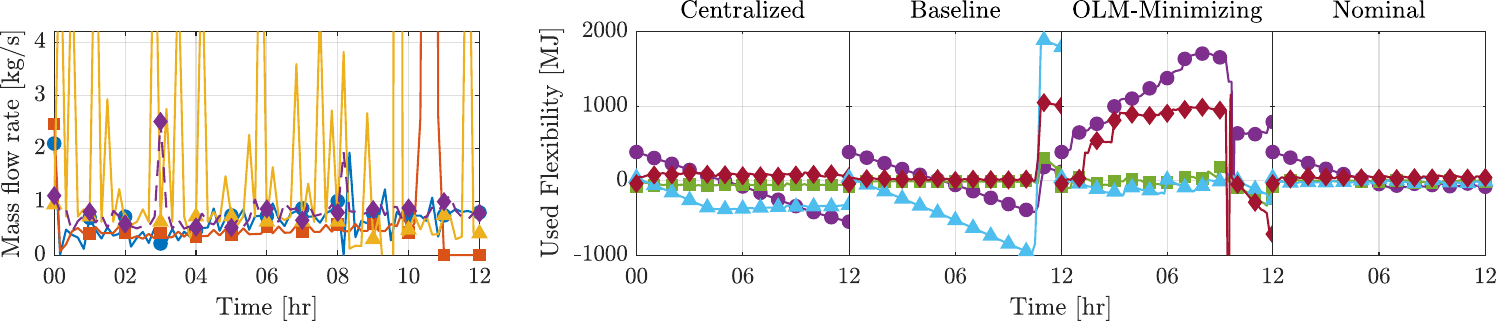}
        \caption{Commercial $U_2$.}
        \label{fig:flex_comm2}
    \end{subfigure}
    \caption{Plant mass flow rate and used flexibility for each of the 13 cases considered. Shows results for a centralized controller, and distributed controllers designed using each of the three partitions: the baseline method, the updated OLM-minimizing partition, and the partition found for the nominal case.}
    \label{fig:flex_all}
\end{figure}
\end{document}